\documentclass[a4paper,11pt]{article} 
\pdfoutput=1

\usepackage{graphicx,rotating,hyperref,slashed,amsmath,xcolor,amssymb,amsfonts,colortbl,cite,soul,cancel} 
\usepackage[export]{adjustbox}
\makeatletter   
\hypersetup{colorlinks,bookmarksopen,bookmarksnumbered,
linkcolor=blue,pdfstartview=FitH,urlcolor=rossos,citecolor=verde}
\allowdisplaybreaks	 
\numberwithin{equation}{section} 
  
\hypersetup{%
    ,urlcolor=black
    ,citecolor=black
    ,linkcolor=black 
    }
\usepackage{mciteplus}

\AtBeginDocument{
  \hypersetup{
    urlcolor=blue,
    citecolor=blue,
    linkcolor=blue,
  }%
}

\newcommand{\hv}[1]{\ensuremath{\hat{#1}}}
\newcommand{\hn}{\ensuremath{\hat{n}}}

\newcommand{\qev}[1]{\ensuremath{\langle {#1} \rangle}}
\newcommand{\be}{\begin{equation}}
	\newcommand{\ee}{\end{equation}}
\newcommand{\bea}{\begin{eqnarray}}
	\newcommand{\eea}{\end{eqnarray}}
\newcommand{\barr}{\begin{array}}
	\newcommand{\earr}{\end{array}}
\newcommand{\ba}{\begin{align}}
	\newcommand{\ea}{\end{align}}

\newcommand{\Omegagw}{\ensuremath{\Omega_{\rm GW}}}

\newcommand{\lc}{\ensuremath{\mathcal{L}}}
\newcommand{\oc}{\ensuremath{\mathcal{O}}}
\newcommand{\rc}{\ensuremath{\mathcal{R}}}
\newcommand{\fc}{\ensuremath{\mathcal{F}}}
\newcommand{\pc}{\ensuremath{\mathcal{P}}}

\newcommand{\SNR}{\text{SNR}}

\def\bea{\begin{eqnarray}}
\def\eea{\end{eqnarray}} 

\def\be{\begin{equation}}
\def\ee{\end{equation}} 
\def\beq{\begin{equation}}
\def\eeq{\end{equation}}

\newcommand\ees{\end{eqnarray}}
\newcommand\bees{\begin{eqnarray}}

\def\bea{\begin{eqnarray}}
\def\eea{\end{eqnarray}}

\def\0{{\boldsymbol 0}}

\def\lsim{\mathrel{\rlap{\lower3pt\hbox{\hskip0pt$\sim$}}
   \raise1pt\hbox{$<$}}}         
\def\gsim{\mathrel{\rlap{\lower4pt\hbox{\hskip1pt$\sim$}}
   \raise1pt\hbox{$>$}}}         

 \newcommand{\sfootnote}[1]{}
\definecolor{bluc}{cmyk}{1,1,0,0.1}
\definecolor{rossoCP3}{cmyk}{0,.88,.77,.40}
\definecolor{rosso}{cmyk}{0,1,1,0.4}
\definecolor{rossos}{cmyk}{0,1,1,0.55}
\definecolor{rossoc}{cmyk}{0,1,1,0.2}
\definecolor{verdes}{cmyk}{0.92,0,0.59,0.4}

\hypersetup{colorlinks, bookmarksopen, bookmarksnumbered,
citecolor=verdes, linkcolor=bluc, pdfstartview=FitH, urlcolor=rossos}

\usepackage{multicol}
\usepackage{color}
\definecolor{rosso}{cmyk}{0,1,1,0.4}
\definecolor{rossos}{cmyk}{0,1,1,0.55}
\definecolor{rossoc}{cmyk}{0,1,1,0.2}
\definecolor{blu}{cmyk}{1,1,0,0.3}
\definecolor{blus}{cmyk}{1,1,0,0.6}
\definecolor{bluc}{cmyk}{1,1,0,0.1}
\definecolor{verde}{cmyk}{0.92,0,0.59,0.25}
\definecolor{verdec}{cmyk}{0.92,0,0.59,0.15}
\definecolor{verdes}{cmyk}{0.92,0,0.59,0.4}

\skip\@mpfootins = \skip\footins
\renewcommand\&{&}

\def\circa#1{\,\raise.3ex\hbox{$#1$\kern-.75em\lower1ex\hbox{$\sim$}}\,}

\newfam\rsfsfam
\def\mathscr#1{{\fam\rsfsfam\relax#1}}

\def\circa#1{\,\raise.3ex\hbox{$#1$\kern-.75em\lower1ex\hbox{$\sim$}}\,}
\makeatletter

\def\hhref#1{\href{http://arxiv.org/abs/#1}{arXiv:#1}} 

\newcommand{\doi}[1]{\href{http://dx.doi.org/#1}{[doi]}}

\def\hhref#1{\href{http://arxiv.org/abs/#1}{arXiv:#1}} 
 
\def\art{\@ifnextchar[{\eart}{\oart}}
\def\eart[#1]#2#3#4#5#6{{\rm #2}, {\em #3 \bf #4} {\rm (#6) #5} ({\em #1})}

\def\article{\@ifnextchar[{\earticle}{\oarticle}}
\def\oarticle#1#2#3#4#5#6{{\rm #1}, {\em ``#6''}, {\rm #2 #3 (#5) #4}}
\def\earticle[#1]#2#3#4#5#6#7{{\rm #2}, {\em ``#7''}, {\rm #3 #4 (#6) #5}  [\hhref{#1}]}
\def\hepart[#1]#2{{\rm #2, \em#1}}
\def\heparticle[#1]#2#3{#2, {\em ``#3''} [\hhref{#1}]}

\newcounter{alphaequation}[equation]
\def\thealphaequation{\theequation\hbox to
0.6em{\hfil\alph{alphaequation}\hfil}}
\def\eqnsystem#1{
\def\@eqnnum{{\rm (\thealphaequation)}}
\def\@@eqncr{\let\@tempa\relax \ifcase\@eqcnt \def\@tempa{& & &} \or
  \def\@tempa{& &}\or \def\@tempa{&}\fi\@tempa
  \if@eqnsw\@eqnnum\refstepcounter{alphaequation}\fi
\global\@eqnswtrue\global\@eqcnt=0\cr}
\refstepcounter{equation} \let\@currentlabel\theequation \def\@tempb{#1}
\ifx\@tempb\empty\else\label{#1}\fi
\refstepcounter{alphaequation}
\let\@currentlabel\thealphaequation
\global\@eqnswtrue\global\@eqcnt=0 \tabskip\@centering\let\\=\@eqncr
$$\halign to \displaywidth\bgroup \@eqnsel\hskip\@centering
$\displaystyle\tabskip\z@{##}$&\global\@eqcnt\@ne
\hskip2\arraycolsep\hfil${##}$\hfil& \global\@eqcnt\tw@\hskip2\arraycolsep
$\displaystyle\tabskip\z@{##}$\hfil
\tabskip\@centering&\llap{##}\tabskip\z@\cr}
\def\endeqnsystem{\@@eqncr\egroup$$\global\@ignoretrue} \makeatother

\definecolor{fiorentina}{rgb}{.5,0,.5}

\begin{document}

\setcounter{page}{1} \baselineskip=15.5pt \thispagestyle{empty}

\vspace{0.8cm}
\begin{center}

{\fontsize{19}{28}\selectfont  \sffamily \bfseries {Measuring kinematic anisotropies \\
\vskip0.15cm
with 
pulsar timing arrays
}}

\end{center}

\vspace{0.2cm}

\begin{center}
{\fontsize{13}{30}\selectfont  
N.~M.~Jim\'enez Cruz$^{1}$, Ameek Malhotra$^{1}$, Gianmassimo Tasinato$^{1,2}$, Ivonne Zavala$^{1}$
} 
\end{center}

\begin{center}

\vskip 8pt
\textsl{$^{1}$ Physics Department, Swansea University, SA28PP, United Kingdom}\\
\textsl{$^{2}$ Dipartimento di Fisica e Astronomia, Universit\`a di Bologna,\\
 INFN, Sezione di Bologna, I.S. FLAG, viale B. Pichat 6/2, 40127 Bologna,   Italy}
\vskip 7pt

\end{center}

\smallskip
\begin{abstract}
\noindent
Recent Pulsar Timing Array (PTA) collaborations show strong evidence for a stochastic gravitational wave background (SGWB) with the characteristic Hellings-Downs inter-pulsar correlations. The signal may stem from supermassive black hole binary mergers, or early universe phenomena. The former is expected to be strongly anisotropic while primordial backgrounds are likely to be predominantly isotropic with small  fluctuations. In case the observed SGWB is of cosmological origin, our relative motion with respect to the SGWB rest frame is a guaranteed source of anisotropy, leading to $\mathcal{O}(10^{-3})$ energy density fluctuations of the SGWB. {For such cosmological SGWB,}  kinematic anisotropies are likely to be larger than the intrinsic anisotropies, akin to the cosmic microwave background (CMB) dipole anisotropy. 
We assess the sensitivity of current PTA data to the kinematic dipole anisotropy, and we also  forecast at what extent
  the magnitude and direction of the kinematic dipole can  be measured in the future with an SKA-like experiment. We also discuss how the spectral shape of the SGWB and the location of the pulsars to monitor affect the prospects of detecting the kinematic dipole with PTA. In
the future, a detection of this anisotropy may even help resolve the discrepancy in the magnitude of the kinematic dipole as measured by CMB and large-scale structure observations.
\end{abstract}

\section{\texorpdfstring{Introduction} {Introduction}}
\label{sec_intr}

Various Pulsar Timing Array (PTA) collaborations recently reported the detection of a stochastic gravitational wave background (SGWB) of unknown origin~\cite{NANOGrav:2023gor,Reardon:2023gzh,Xu:2023wog,EPTA:2023fyk,InternationalPulsarTimingArray:2023mzf}. Determining the sources of the SGWB --  whether it is generated from the mergers of supermassive black hole binaries~\cite{Sesana:2008mz,Burke-Spolaor:2018bvk}, or via high energy processes in the early universe (see e.g.~\cite{Caprini:2018mtu} for a review) -- requires the detection of SGWB properties beyond the spectral shape and amplitude of the SGWB intensity. The anisotropy of the SGWB is one such key property that can  discriminate among an astrophysical or cosmological origin of the observed signal. On the one hand, a strong level of anisotropy is expected for an astrophysical signal, due to the clustering of galaxies where the GW sources reside, as well as Poisson-type fluctuations in the source properties~\cite{Cornish:2013aba,Mingarelli:2013dsa,Taylor:2013esa,Mingarelli:2017PTA,Cornish:2015ikx,Taylor:2020zpk,Becsy:2022pnr,Allen:2022dzg,Sato-Polito:2021efu,
Sato-Polito:2023spo}. On the other hand, cosmological sources are expected to have a small ($\sim 10^{-5}$) level of intrinsic anisotropy (see e.g. \cite{Alba:2015cms,Contaldi:2016koz,Geller:2018mwu,Bartolo:2019oiq,Bartolo:2019zvb,Bartolo:2019yeu}, and  \cite{LISACosmologyWorkingGroup:2022kbp} for a review). However, in addition to  intrinsic anisotropies, cosmological SGWB are  expected to be characterised by a level of kinematic Doppler anisotropy similar to what has been observed by cosmic microwave background (CMB) experiments. According to CMB observations, our velocity with respect to the cosmic rest frame has a magnitude $\beta = v/c\,=\,1.23\times 10^{-3}$,  and is directed towards $(l,b)=(264^{\circ},48^{\circ})$ in galactic coordinates \cite{Smoot:1977bs,Kogut:1993ag,WMAP:2003ivt,WMAP:2008ydk,Planck:2013kqc}. 

The capability of PTAs to detect SGWB anisotropy has been previously studied in~\cite{Anholm:2008wy,Taylor:2013esa,Mingarelli:2013dsa,Gair:2014rwa,Cornish:2014rva,Mingarelli:2017fbe,Hotinli:2019tpc,Ali-Haimoud:2020ozu,Ali-Haimoud:2020iyz,Pol:2022sjn}, also applying and extending techniques developed in the context
of ground-based  and space-based experiments, see e.g. \cite{Allen:1996gp,Ballmer:2005uw,Thrane:2009fp,Renzini:2018vkx,Payne:2020pmc,Cornish:2001hg,Baker:2019ync,Banagiri:2021ovv,Contaldi:2020rht,Mentasti:2020yyd,Mentasti:2023icu}.
 Although  the SGWB has not yet been detected with  sufficient sensitivity to allow for the detection of its anisotropies~\cite{Taylor:2015udp,NANOGrav:2023tcn},  longer observation times,  as well as the addition of more pulsars and a better
determination of their intrinsic 
noise  \cite{Chu:2021krj,Bernardo:2022rif,Bernardo:2023mxc},  
might change the situation. This will be particularly true in the SKA era:
see for example \cite{Janssen:2014dka} for a general analysis of prospects of GW detections based on the SKA  experiment. If the observed signal is indeed cosmological, it is conceivable that the Doppler anisotropy will be the next fundamental  property of the SGWB to be measured. The dipole of this anisotropy, henceforth referred to as the `kinematic dipole',  will be the main focus of this paper. It has an amplitude of the order $\beta$, and is also strongly dependent on the spectral shape of the SGWB. The PTA response to the dipole has also been previously investigated in~\cite{Anholm:2008wy,Mingarelli:2013dsa,Tasinato:2023zcg} and a significant dependence on the locations of the pulsars being observed has been found. Thus, the detection of the kinematic dipole depends crucially on all these different parameters,  making it important to investigate how best to maximise PTA sensitivity to this specific SGWB anisotropy.  Notice that  kinematic anisotropies are also a topic of active research in the context of GW interferometers \cite{LISACosmologyWorkingGroup:2022kbp,Cusin:2022cbb,Bertacca:2019fnt,ValbusaDallArmi:2022htu,Chung:2022xhv,Chowdhury:2022pnv}.

Besides  helping us understand the origin of the SGWB, a detection of the kinematic dipole with PTA may well serve another purpose by providing us an independent measurement of our velocity with respect the cosmic rest frame.\footnote{It is reasonable to assume that the SGWB and CMB rest frames coincide, unless there is significant CMB-GW isocurvature in the early universe. See e.g. \cite{Turner:1991dn,Langlois:1995ca,Domenech:2022mvt} for studies exploring this possibility. Other options include deviations from the cosmological principle, see e.g. \cite{Aluri:2022hzs} for a review.
}
 Another way to measure the kinematic dipole is through Large Scale Structure (LSS) observations --- using the number counts of radio sources and of quasars. This method has recently come in tension with the results obtained from the CMB. While the CMB and LSS results broadly agree on the direction of our motion, they differ significantly on the magnitude of the velocity. See e.g. \cite{Colin:2017juj,Bengaly:2017slg,Siewert:2020krp,Secrest:2020has,Dalang:2021ruy,Secrest:2022uvx,Dam:2022wwh,Wagenveld:2023kvi,Singal:2023wni}, as well as \cite{Peebles:2022akh} for a critical assessment. It is thus also interesting to explore if GW observations can provide insights related to this discrepancy. Last, but not least, information on kinematic anisotropies with PTA measurements can be used to set constraints on alternative theories of gravity, see e.g. \cite{Tasinato:2023zcg,AnilKumar:2023yfw}.
 
 Motivated by these facts,
in this paper we investigate the current and future sensitivity of PTAs to the SGWB kinematic dipole, {focusing on cosmological SGWB where this dipole anisotropy is expected to be dominant}. We begin in section \ref{sec_basicf} by reviewing the generation of kinematic anisotropies and the PTA response to them. In section \ref{sec_pres}, we estimate the sensitivity of the current-type PTA datasets to the kinematic dipole, and
in section \ref{sec:Fisher_ideal} we provide Fisher forecasts for its measurement with a futuristic SKA-type PTA experiment. We focus on the parameters associated to the SGWB intensity (the amplitude $A$ and the spectral shape $\gamma$) and the kinematic dipole ($\beta$, $\hat v$), highlighting the role played the SGWB spectral shape and locations of pulsars in the determination of these quantities. We also estimate the PTA sensitivity required for a measurement of $\beta$ at a level precise enough to shed light on the CMB-LSS dipole tension discussed above.
We will learn that  monitoring a large number of pulsars is needed for acquiring sufficient sensitivity, and their positions in the sky might be the key property to detect kinematic effects.
 Finally, we present our conclusions in~\ref{sec:conclusion}, followed
by three technical appendices.

\section{\texorpdfstring{Set-up}{Set-up}}
\label{sec_basicf}

In this  section we briefly review how to describe the kinematic
anisotropies of the {cosmological} SGWB, and how to measure them with PTA experiments. We are particularly
interested in investigating how the PTA sensitivity to Doppler effects depends on the pulsar positions, and on the frequency profile of the intensity  of the SGWB. 
For more details, the reader may consult \cite{Tasinato:2023zcg} and
references therein. GW correspond to
transverse-traceless perturbations of the flat metric. They are controlled by a tensorial quantity $h_{ij}(t,\vec x)$, small in size with respect to the background:
\be
ds^2\,=\,-d t^2+\left(\delta_{ij}+h_{ij}(t,\vec x) \right)\,d x^i d x^j
\,.
\ee
The tensor $h_{ij}(t,\vec x)$ can be decomposed in Fourier modes as (we set
$c=1$)
\be
h_{ij}(t,\vec x) \,=\,\sum_\lambda\,\int_{-\infty}^{\infty}\,d f\,\int d^2 \hat  n \,
e^{2 \pi i f \left(t- \hat n\cdot \hat x \right)}\,{\bf e}_{ij}^{\lambda} (\hat n)\,h_{\lambda}(f, \hat n)\,.
\ee
The quantities ${\bf e}_{ij}^{\lambda} (\hat n)$ are the  polarization
tensors for the two GW helicities $\lambda\,=\,+,\,\times$, which depend
on the GW direction $\hat n$. Besides $\hat n$, the GW Fourier mode $h_{\lambda}(f, \hat n)$ depends also on the GW frequency $f$. We assume that the two point
function among Fourier fields $ h_\lambda (f, \hat n)$ obeys the relation
\be
\label{h2pt}
  \langle h_\lambda (f, \hat n)\, h_{\lambda'} (f', \hat n') \rangle 
  \,=\,\frac12\,\delta_{\lambda \lambda'}\,\delta(f-f')\,\frac{\delta(\hat n-\hat n')}{4\pi}
  \,I(f, \hat n)\,,
  \ee
valid for Gaussian, unpolarised and stationary SGWB. In principle,
the GW intensity $I(f, \hat n)$ 
depends both on frequency and direction of GW. In this work, its  directional
dependence will be associated with Doppler effects. We can define an isotropic
version  of the intensity by averaging over directions
\be
\label{defiba}
\bar I(f)\,=\,\frac{1}{4 \pi}\int d^2 \hat n\,I(f,\hat n)
\,.\ee
$\bar I(f)$ is 
a quantity 
which we  use in what follows. 

\subsubsection*{GW and PTA experiments}

We now briefly review how
PTA experiments respond to GW (see e.g. \cite{Maggiore:2018sht} for more extensive
discussions). We introduce the time delay $z_a(t)$ of light geodesics,
induced by GW crossing the space between
 the
earth and
a given pulsar 
 $a$:
 \be
  z_a(t)\,\equiv\,\frac{\Delta T_a}{T_a}\,=\sum_\lambda 
  \int df d^2 \hat  n \,
  e^{2 \pi i f t}\,{\bf e}_{ij}^{\lambda} \,D_a^{ij}(\hat n)\,h_\lambda(f, \hat n)\,.
  \ee
    $T_a$ is the pulsar period in absence of GW,  and
\be
  D_a^{ij}(\hat n)\,=\,\frac{\hat x_a^i\,\hat x_a^j}{2 (1+\hat n \cdot \hat x_a)}
  \ee
 is the so-called detector tensor. While the earth is located at the origin
 of our coordinate system, the pulsar is located
 at position $\vec x_a\,=\,\tau_a \,\hat x_a$, with $\tau_a$ the time of travel of 
 the pulsar signal from emission to detection. 

Using the previous ingredients, in particular relation \eqref{h2pt},
we can build the two point function for time delays between a pulsar
pair $(a b)$
\bea
\label{defzazb}
  \langle z_a (t_1) z_b (t_2)
  \rangle
  &=&\frac12 \int d f\,d^2 \hat n\,  \gamma_{ab}(\hat n)
\,I(f,\hat n)\,\cos{\left(2 \pi f \,(t_1-t_2) \right)}\,.
  \eea
The tensor $ \gamma_{ab}(\hat n)$ depends on the GW
direction, and is given by
  \bea
  \gamma_{ab}(\hat n)
  &=&
\sum_\lambda   {\bf e}_{ij}^{\lambda} \,D_a^{ij}(\hat n)\,  {\bf e}_{pq}^{\lambda} \,D_b^{pq}(\hat n)
   \nonumber
   \\&=&\frac{(\hat x_a \cdot \hat n)^2+(\hat x_b \cdot \hat n)^2+ (\hat x_a \cdot \hat n)^2  (\hat x_b \cdot \hat n)^2-1}{8(1+\hat x_a \cdot \hat n)(1+\hat x_b \cdot \hat n)}
\nonumber\\
&&+\frac{ (\hat x_a \cdot \hat x_b)^2-2 (\hat x_a \cdot \hat x_b)  (\hat x_a \cdot \hat n)  (\hat x_b \cdot \hat n) 
}{4(1+\hat x_a \cdot \hat n)(1+\hat x_b \cdot \hat n)}
\,.
\label{contra}
  \eea
  We can also explicitly perform the integration over directions in eq \eqref{defzazb}. Defining
  $\Delta t_{12}=t_2-t_1$, we obtain
   {
    \bea
  \langle z_a (t_1) z_b (t_2)
  \rangle
  &=&\frac12 \int d f\,\cos\left(
  2 \pi f \Delta t_{12}
  \right)
  \Gamma_{ab}(f)\,\bar{I}(f)\,,
  \eea  
  }
   where we   use the so-called PTA overlap function
\bea
\label{defgab}
\Gamma_{ab}(f)\,=\,\int d^2 \hat n\,  \gamma_{ab}(\hat n)
\,\frac{I(f,\hat n)}{\bar I(f)}\,.
\eea
So far, our formulas are  general. We will soon specialise
to the case of anisotropies in the GW intensity, as induced by kinematic effects. But first, we introduce  the concept of time residual
\be
R_a(t)\,\equiv\,\int_0^t\,d \tilde t\,z_a(\tilde t)\,,
\ee
more commonly used in analysing PTA experiments. 
Analogously to the time-delay case, we can form its two point correlation function
{ \begin{eqnarray}
 \langle R_a(t_A) R_b(t_B) \rangle
 &=& \frac12 \int_0^{t_A} \int_0^{t_B}
 d t_1\,d t_2\int df d^2 \hat n \,
 \left[
 \cos\left(
  2 \pi f \Delta t_{12}
  \right)
  \gamma_{ab}(\hat n)\,{I}(f,\hat n)
  \right]
 \\
 &=&
 \int df d^2 \hat n \,\frac{\sin (\pi t_A f) \sin (\pi t_B f) }{\pi\,f^2}\, \gamma_{ab}(\hat n)\,
{I(f,\hat n)} \cos\left(
  2 \pi f \Delta t_{AB}
  \right)\,,
\label{r2pt1}
 \end{eqnarray}}
with $\gamma_{ab}$
defined in eq \eqref{contra}. The $\sin$ and
$\cos$ coefficients give order one contributions,
and they are usually neglected in the definition
of the time residual
correlation functions (see e.g. \cite{Maggiore:2018sht}, chapter 23). We consequently parameterize eq \eqref{r2pt1} as 
\cite{Maggiore:2018sht}
\bea
  { R}^{\rm GW}_{ab}
 &=&\int df
 d^2 \hat n\, \gamma_{ab}(\hat n)\,
\frac{I(f,\hat n)}{(4 \pi f)^2}
\label{r2pt2}
 \\&=&
 \int df\,\Gamma_{ab}(f)\,
\frac{\bar I(f)}{(4 \pi f)^2}\,,
\label{r2pt3}
 \eea
 where in passing from line \eqref{r2pt2} to \eqref{r2pt3} we integrate
 over directions, and use the definition \eqref{defgab}. Besides the GW signal
 of eq \eqref{r2pt2} (if present), we expect that undesired  noise sources give contributions $ { R}^{\rm N}_{ab}$
 to the time residual correlation functions. We write  the total signal
 as
 \be
 \label{deftrab}
  {\cal R}_{ab}\,=\,  { R}^{\rm GW}_{ab}+  { R}^{\rm N}_{ab}\,,
 \ee 
 where
\be
 { R}^{\rm N}_{ab}\,=\,\delta_{ab} \sigma_a^2\,,
\ee
and $ \sigma_a^2$ is the variance of the noise sources at each pulsar\footnote{Here we  assume that noise is uncorrelated across pulsars. However, common noise sources do exist, see e.g.
\cite{Tiburzi_2015}.}:
\begin{align}
    \qev {N_a(f)N^*_b(f') } = \delta_{ab}\frac{\delta(f-f')}{2}\sigma^2_a(f)\,.
\end{align}

\subsubsection*{Kinematic anisotropies}
We now specialise to the case of kinematic 
anisotropies of the SGWB. We make use of formulas first developed
in \cite{Cusin:2022cbb}. 
  We assume 
that the GW intensity is intrinsically isotropic: $\bar I(f)$ (it is straightforward
to extend our formulas to the general case, see e.g. \cite{Cusin:2022cbb}). The intensity 
only develops a kinematic anisotropy, due to our motion with respect to the SGWB
rest frame  whose velocity has magnitude $\beta$ and direction $\hat v$ (in our
units with $c=1$).  This assumption might be justified for a cosmological SGWB, while astrophysical
backgrounds are characterized by intrinsic 
anisotropies of size larger than the kinematical ones (see the discussion
in section \ref{sec_intr}).

Using the formulas in \cite{Cusin:2022cbb}, one finds \cite{Tasinato:2023zcg}:
\be
\label{deftri1}
\frac{I(f,\hat n)}{\bar I(f)}\,=\,{\cal D}\,\frac{\bar I\left ({\cal D}^{-1}f\right)}{\bar I(f)}
\,,
\ee
where the quantity
\be
{\cal D}\,=\,\frac{\sqrt{1-\beta^2}}{1-\beta\, \hat n \cdot \hat v}
\ee
is a kinematic factor, depending on the angle between the GW direction
$\hat n$ and the relative velocity $\hat v$ among frames.
The structure of \eqref{deftri1}
makes manifest that kinematic  anisotropies are {\it not} factorisable in a part depending on frequency, times a part depending on direction.
 Kinematic anisotropies are factorizable
 only for intensities with a power-law frequency dependence profile \cite{Cusin:2022cbb,Chowdhury:2022pnv}. 

\smallskip

CMB experiments  indicate that the typical size of $\beta$
is small, of order one per mil (see the discussion
in section \ref{sec_intr}). Hence, we can Taylor
expand \eqref{deftri1} in terms of the small parameter $\beta$, and   keep
only the first terms in the expansion. In this work, we mainly focus on the leading contribution
of order $\beta$, corresponding to  dipolar anisotropies. But, in principle, we can also consider the next-to-leading contribution of order $\beta^2$, associated with
  quadrupolar Doppler effects.  For this reason, we develop our formulas up to order $\beta^2$. Taylor expanding eq \eqref{deftri1}, we find
\bea
\frac{I(f,\hat n)}{\bar I(f)}&=&\left[
1-\frac{\beta^2}{6} \left(1-n_I^2-\alpha_I \right)\right]
+\beta\,\hat n\cdot \hat v\, \left(1-n_I \right)
\nonumber
\\
&&
+\frac{\beta^2}{2} \left( (\hat n\cdot \hat v)^2-\frac13
 \right)\left(2-3 n_I +n_I^2+\alpha_I\right)+{\cal O}(\beta^3)\,.
 \label{tayexa}
\eea
The tilt parameters are defined as 
\be
\label{deftil}
 n_I\,=\,\frac{d\,\ln \bar I}{d\,\ln f}\hskip0.5cm,\hskip0.5cm \alpha_I\,=\,\frac{d\,n_I}{d\,\ln f}\,.
 \ee
Notice that,  in general, these quantities depend on frequency. Plugging
expression \eqref{tayexa} into eq \eqref{defgab}, we can straightforwardly perform
the angular integrations \cite{Tasinato:2023zcg}, and obtain the following expression
\bea
\label{respia}
\Gamma_{ab}(f)&=&\left[
1-\frac{\beta^2}{6} \left(1-n_I^2-\alpha_I \right)\right]\,\Gamma^{(0)}_{ab}+
\beta\,\left(n_I -1\right)  \Gamma_{ab}^{(1)}
\nonumber
\\
&&
+\frac{\beta^2}{2}\left(2-3 n_I +n_I^2+\alpha_I\right)\,\Gamma^{(2)}_{ab}\,,
\eea
for the PTA overlap function among a pulsar pair $(ab)$. The quantities   $\Gamma_{ab}^{(i)}$ are given by \cite{Tasinato:2023zcg}
\bea
\label{defFHD}
\Gamma^{(0)}_{ab}
&=&\frac13-\frac{y_{ab}}{6}+y_{ab} \ln y_{ab} 
\\
\Gamma^{(1)}_{ab}&=& 
\left(\frac{1}{12}+\frac{ y_{ab}}{2}+\frac{ y_{ab} \ln y_{ab}}{2(1-y_{ab})} \right)
\, \left[\hat v\cdot \hat x_a+\hat v\cdot \hat x_b\right]\,,
\label{defF11}
\\
\Gamma^{(2)}_{ab}&=&
\left(\frac{3-13  y_{ab}}{20 ( y_{ab}-1)}+\frac{ y_{ab}^2 \ln  y_{ab}}{2(1- y_{ab})^2} \right)
\,\left[(\hat v\cdot \hat x_a)(\hat v\cdot \hat x_b) \right]
\nonumber\\
&+&
\left( \frac{1+2  y_{ab}-4  y_{ab}^2+ y_{ab}^3+3  y_{ab} \ln  y_{ab}
}{
12 (1- y_{ab})^2
}\right)\,\left[(\hat v\cdot  \hat x_a)^2+(\hat v\cdot  \hat x_b)^2 \right]
\,,
\label{defF12}
\eea
with
\be\label{defyab}
{y}_{ab}\,=\,\frac{1-\hat x_a  \cdot \hat x_b}{2}\,=\,\frac{1-\cos{\zeta_{ab}}}{2}\,,
\ee
and $\zeta_{ab}$ the angle between the two pulsar vectors $\hat x_a$ and $\hat
x_b$. 
 
\begin{figure}[t]
	\centering
	\includegraphics[width=0.45\linewidth]{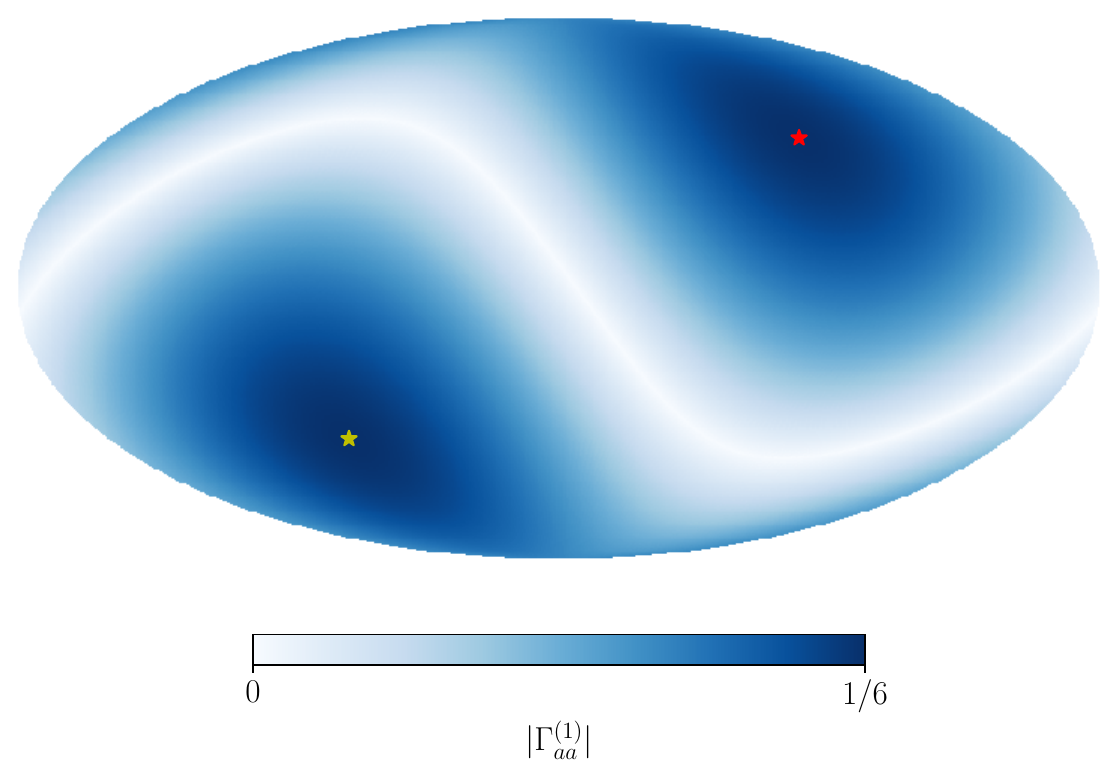}
 	\includegraphics[width=0.45\linewidth]{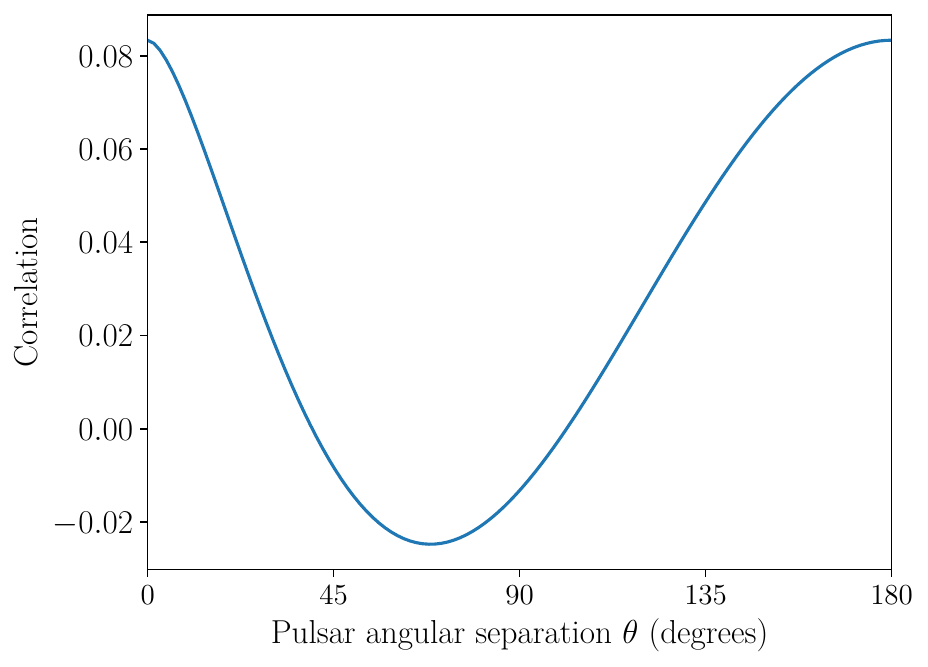}
	\caption{  \it{Left: the magnitude of the PTA response function  $\Gamma^{(1)}_{ab}$  to kinematic dipole anisotropies, as defined in eq \eqref{defF11}. We fix the velocity vector $\hat v$ along the the direction measured by the CMB ($\hat v$ and $-\hat v$ are denoted by red and yellow stars respectively). We plot the response as a function of the positions
of a pair of pulsars, for simplicity oriented in the same direction. 
    Right: the dipole response as a function of the angle between the pulsars, without  including the $(\hat v\cdot \hat x_a+\hat v\cdot \hat x_b)$ factor in eq \eqref{defF11}.}}
	\label{fig:response_I}
\end{figure}

The quantity $\Gamma^{(0)}_{ab}$ is the well-known Hellings Downs overlap function
\cite{Hellings:1983} (see also the recent \cite{Romano:2023zhb,Kehagias:2024plp} for nice discussions and new perspectives on its physical properties). Kinematic effects modulate the  Hellings Downs curve through a frequency-dependent coefficient, starting
at order $\beta^2$ in our expansion. Moreover, they  add new dipolar  (at order $\beta$)
and quadrupolar (at order $\beta^2$) 
contributions  to the PTA response function. See e.g. Fig \ref{fig:response_I}
for a graphical representation of the quantity $\Gamma^{(1)}_{ab}$.

We stress
two important properties
of the results reviewed so far
\begin{enumerate}
\item
The quantities $\Gamma^{(1,2)}_{ab}$ vanish
for pulsars with vectors orthogonal to the velocity: see eqs \eqref{defF11} and \eqref{defF12}. Hence, the magnitude of 
kinematic anisotropies depend  on the
geometric configuration of the pulsar set: pulsars located along the direction
of the velocity vector $\hat v$ are more sensitive to kinematic effects.
\item The effect of anisotropies depend on the frequency profile
of the GW intensity: if $\bar I(f)$  
has features in frequency which amplify the spectral tilts, 
 anisotropy effects can get enhanced as they multiplied by combinations of $n_I$ and $\alpha_I$. 
 In fact, the study of Doppler anisotropies provides additional tools for probing the frequency 
 dependence of the GW intensity, besides more direct reconstruction techniques (see e.g. \cite{Caprini:2019pxz} and
 references therein).
\end{enumerate}
The formulas  presented in this section constitute  the theoretical
background we need next for analysing to what extent PTA experiments
can probe kinematic anisotropies in the SGWB.

\section{Kinematic anisotropies and  present day PTA data}
\label{sec_pres}

We are interested in measuring kinematic Doppler anisotropies in the SGWB with 
the present generation of PTA measurements, starting from the theoretical formulas discussed in section \ref{sec_basicf}.

 For the case of astrophysics backgrounds, the size
of the intrinsic anisotropies is larger than the kinematic ones we are focussing in. For 
cosmological SGWB, though, intrinsic anisotropies are usually  small in amplitude, and kinematic effects  provide the main contribution 
to anisotropies (see our discussion in section \ref{sec_intr}). GW experiments, if able to detect
kinematic anisotropies, offer an independent tool to measure our motion relative
to the cosmic source of GW, possibly helping to address the current anomaly in the measurements 
of the size of the relative velocity $\beta$ among frames  reviewed in section  \ref{sec_intr}.  Additionally, they can provide  independent
tests for alternative theories of gravity \cite{Tasinato:2023zcg,Bernardo:2023mxc}.
 For this reason,  it is worth  exploring
at what extent Doppler effects can be detected with PTA experiments.

In this section, we focus on  current generation PTA experiments,  which monitor  a finite number of pulsars (around one hundred), placed in specific positions in the sky.  We consider this kind of experimental set-up as `realistic'. So far, SGWB anisotropies have not been detected with PTA data: only upper limits on their size have been placed~\cite{NANOGrav:2023tcn}. Using an approach based on the Fisher formalism,  we develop tools for quantitatively searching for Doppler
anisotropies, and for investigating to what extent realistic current-type PTA experiments should be improved towards this aim.\footnote{In section \ref{sec:Fisher_ideal} we  will further develop the topic considering idealized, futuristic experiments monitoring very large numbers of pulsars isotropically distributed in the sky. We then show that the prospects of detection of Doppler anisotropies improve  considerably.}
 In particular, we  consider  the current NANOGrav data set, (hereafter
 NG15) \cite{NANOGrav:2023gor}, and we then proceed demonstrating how including extra pulsars at specific positions
 in the sky  can improve the sensitivity to Doppler
 effects. In fact, we are interested in developing the first of
   our comments at the end of section \ref{sec_basicf}.

\begin{figure}[t!]
    \centering
    \includegraphics[width=0.6\linewidth]{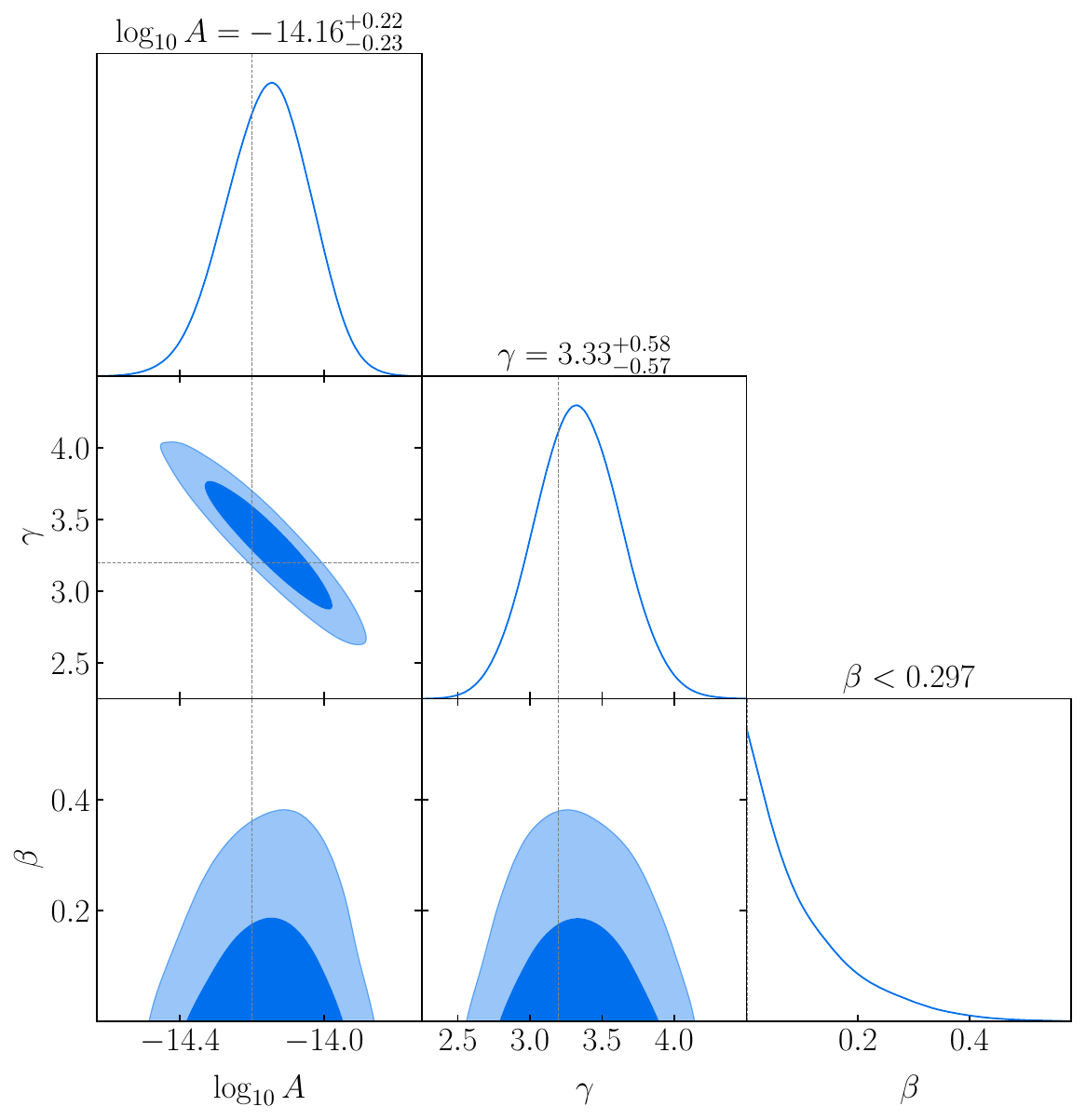}
    \caption{ \it Parameter distributions and $95\%$~C.L limits for the SGWB amplitude $A$ \eqref{I0equivalenceA}, spectral index $\gamma$ \eqref{gammadef} and dipole magnitude~$\beta$. (See main text and Appendix \ref{app_conv} for the definition of these quantities.) The recovered amplitude and tilt are consistent with the NG15 results {(grey markers)}. We also obtain {$\beta < 0.297$} at $95\%$~C.L.}
    \label{fig:NG15_upper limts}
\end{figure}

But before forecasting the capabilities of  PTA experiments  to measure kinematic effects
by  developing a  dedicated Fisher formalism, we start 
by using the  {\it most recent  NG15 data} to extract what information  they can provide on Doppler
effects. Namely,  
we  analyse for the first time the current NG15 dataset \cite{NANOGrav:2023gor}, searching for the presence of the kinematic dipole, fixing the dipole direction to  the one  measured by CMB experiments.\footnote{We have checked that varying the  dipole direction parameters does not provide any meaningful constraints on the dipole direction. This is to be expected given that the isotropic background is only detected with an $\SNR\sim 5$ at present.} The parameters that we vary are 
 $\{\log_{10}A,\gamma,\beta\}$  (see appendix~\ref{app_conv} for our conventions on the quantities characterising the SGWB) and {we fix the red noise parameters (amplitude and spectral index) for each pulsar to their median values from the NG15 analyses for ease of computation}. {Varying the red noise parameters as well will lead to slightly weaker constraints on $\beta$ but the results are not going to be significantly different.} We access the likelihood used by the NANOGrav collaboration through \texttt{ENTERPRISE}~\cite{enterprise} and \texttt{ENTERPRISE\char`_EXTENSIONS}~\cite{enterprise-ext}. The parameter space is explored using a Markov Chain Monte Carlo sampler~\cite{Lewis:2002ah,Lewis:2013hha}, through its interface with \texttt{Cobaya}~\cite{Torrado:2020dgo}. Our results are plotted in Fig \ref{fig:NG15_upper limts} using \texttt{GetDist}~\cite{Lewis:2019xzd}. The parameter limits for the SGWB amplitude and spectral shape are consistent with those obtained in~\cite{NANOGrav:2023gor} and we set an upper limit {$\beta < 0.297$} at $95\%$~C.L, assuming a cosmological origin for the SGWB. Clearly, current data do not reach the per mil sensitivities on $\beta$, which are needed to set meaningful constraints on this quantity. {The upper limit we obtain is consistent with the results of~\cite{NANOGrav:2023tcn}, which shows that the current sensitivity to SGWB anisotropy is limited to about $10\%$ (or larger) relative to the isotropic monopole. Furthermore, we stress that the upper limit we consider should be understood to apply only if the SGWB is of cosmological origin, a case in which the kinematic dipole is expected to be the largest anisotropy by far. Otherwise the intrinsic anisotropy of the SMBHB scenario must also be accounted for in the analyses.} 

\subsection{Extracting information from data}
\label{sec_info}

 We now develop our own application of the Fisher formalism, tailored to extract information on Doppler effects
 from PTA results. We start 
  with a general discussion on how to extract information
 on model parameters from data.
 We find it convenient to adopt the approach and notation from 
 \cite{Ali-Haimoud:2020ozu,Ali-Haimoud:2020iyz}. We denote a given pulsar 
  pair $(a,b)$ with the capital letter $A$: $(a,b)\to A$. The GW-induced 
    time residual  is
  \be
  \label{abRa}
  R_A^{\rm GW}\,=\,\frac{\gamma_A \cdot  I }{(4 \pi f)^2}\,=\,\frac{\Gamma_A \cdot \bar I }{(4 \pi f)^2}
  \,.
  \ee
   The dot $\cdot$ here indicates that we are working 
  with bandwidth-integrated quantities, centered at frequency $f$, that is:
  \be
  \frac{\Gamma_A \cdot \bar I}{(4 \pi f)^2}
\,=\,\int_{f-\Delta f/2}^{f+\Delta f/2} d\tilde f
\,\frac{\Gamma_A(\tilde f)  \bar I(\tilde f)}{(4 \pi\tilde f)^2}\,,
  \ee
 where $ \bar I(f)$ is the isotropic value of the frequency
 given in eq  \eqref{defiba}, and $\Gamma_A=\Gamma_{ab}$ is
 the overlap reduction function of eq \eqref{respia}. 
  We include
  kinematic effects up to the quadrupole
 of order $\beta^2$.  In the equalities of eq \eqref{abRa}, we can
  use $\gamma_A$ or $\Gamma_A$, following the steps between
 eqs \eqref{defzazb} and \eqref{defgab}, and integrating over all directions
 for passing from one equality to the other.  In fact, in the first equality
 of eq \eqref{abRa}, the dot also includes integration over directions.

\smallskip
   
We approximate the likelihood as Gaussian  in the timing residual cross-spectra~\cite{Ali-Haimoud:2020ozu}. We  
 can then write
\bea
    \label{eq:sgwb_likelihood}
    -2\ln \lc &=& {\mathrm{const.}} + \sum_{f}\sum_{AB} \left(\hat{\rc}_A - \frac{\gamma_{A}\cdot  I}{(4\pi f)^2}\right)\,C^{-1}_{AB}\,\left(\hat{\rc}_B - \frac{\gamma_{B}\cdot  I}{(4\pi f)^2 }\right)
\\
 &=& {\mathrm{const.}} + \sum_{f}\sum_{AB} \left(\hat{\rc}_A - \frac{\Gamma_{A}\cdot \bar I}{(4\pi f)^2}\right)\,C^{-1}_{AB}\,\left(\hat{\rc}_B - \frac{\Gamma_{B}\cdot \bar I}{(4\pi f)^2 }\right)
        \label{eq:sgwb_likelihoodV2}
        \,,
\eea
where $\hat{\rc}$ indicate the measured
 cross-correlated timing residuals,
 and $C^{-1}$ is the inverse covariance matrix. In passing from line \eqref{eq:sgwb_likelihood} to \eqref{eq:sgwb_likelihoodV2}, 
 we integrate over directions. 
 As derived in~\cite{Ali-Haimoud:2020ozu}, the covariance matrix of the timing residual band-powers 
 is a combination of time-residual correlators $\rc_{A}\,=\,\rc_{ab}$ 
  with the following structure 
\begin{align}
	\label{eq:cov_def}
    C_{AA'} &
    = \frac{1}{2 T_{AA'} \Delta f}(\rc_{a a'}\rc_{b b'} + \rc_{ab'}\rc_{a'b})\,, 
\end{align}
with $T_{AA'}$ the effective total time of observations of the four pulsars $(a\,b\, a'\, b')$, and we assume that the bandwidth $\Delta f$ satisfies $ 1/T \ll \Delta f \ll f$.
  Furthermore, in the weak signal limit 
 \be
  I \ll (4\pi f^2 \sigma^2)\,,
  \ee
  which we assume from now on,
   we have  (recall the definition of ${\cal R}_A$ in eq \eqref{deftrab})
\begin{align}
    \rc_{A} \approx R_{A}^N\,=\,
    \sigma^2_a\,,
\end{align}
where $\sigma^2_a$ is the (band-integrated) variance of the noise in
 pulsar $a$. 
Thus, in this limit, the diagonal elements are much larger than the off-diagonal ones. Therefore, the covariance can be effectively approximated as diagonal in the pulsar pairs $A,A'$~\cite{Ali-Haimoud:2020ozu} (see also~\cite{Romano:2016dpx} for the same approximation in the context of interferometers, as well as our appendix \ref{appcov}). 
\begin{align}
    \label{eq:cov_weaksignal}
    C_{AA'} \approx \frac{R_{A}^N\,R_{A'}^N}{2 \,T_{AA'}\Delta f}\delta_{AA'}\,,
\end{align}
 Its inverse is also diagonal
\begin{align}
    (C)^{-1}_{AA'} = \frac{2 \,T_{AA'} \Delta f}{R_{A}^N\,R_{A'}^N}\delta_{AA'}\,.
\end{align}
This property
renders our analysis particularly straightforward.  Notice that,  in this approximation, the
covariance matrix depends only on noise parameters.

\smallskip
From the likelihood \eqref{eq:sgwb_likelihoodV2} we can determine the best-fit values of parameters, and their errors. We assume that the combinations $\Gamma\cdot \bar I$ appearing
in the likelihood of eq \eqref{eq:sgwb_likelihoodV2} 
depend on a vector  $\vec \Theta$, whose components correspond
to the  parameters we wish to determine. For example,  the behaviour
of the intensity $\bar I(f)$ as function of frequency. Or,  the size of our velocity $\beta$ with respect 
to the SGWB rest frame. The best fit values for the parameters are found extremising the function
$ \ln {\cal L}$:
\be
\frac{\partial\, {\ln {\cal L}}}{\partial \Theta_i}\,=\,0\,.
\ee 
The errors on the determination of parameters are controlled by
the Fisher matrix,  defined as~\cite{Tegmark:1996bz} (see also \cite{Dodelson:2003ft} for a pedagogical account)
\begin{align}
\label{deffim}
        \fc_{ij} = \left\langle -\frac{\partial \ln \lc}{\partial \Theta_i \partial \Theta_j} \right \rangle\,,
    \end{align}
    in terms of the expectation value of the second derivatives of $ \ln {\cal L}$.
The minimum error on the measurement of the parameters $\Theta_i$ is (no sum on repeated indexes):
    \begin{align}
        \Delta\Theta_i = \sqrt{({\cal F}^{-1})_{ii}}\,,
    \end{align}
giving us an estimate of a lower bound on the experimental errors
 to measure our signal parameters.
   
\subsection{Best-fit values for the model parameters: an example}

We already emphasised that  the PTA response to Doppler
anisotropies depend on the relative position of the monitored pulsars
with respect to the direction of the velocity vector among frames, see section \ref{sec_basicf}. We now discuss how to practically exploit this property. We focus in this section on the specific case of { power law} frequency profile
in the SGWB intensity  rest frame:
 \be
 \label{defplf}
\bar I(f)\,=\, I_0\,\left({f}/{f_\star} \right)^{n_I}\,,
\ee
with $I_0$ a constant amplitude, $n_I$ the constant spectral tilt, and $f_\star$ a reference
frequency. By extremising the log likelihood along the parameter models, see eq \eqref{eq:sgwb_likelihoodV2}, we
 discuss how combinations of measurements allow us to determine the values of $I_0$ -- the amplitude of GW intensity -- and $\beta$ -- the magnitude of the relative velocity among frames.

We vary $ \ln {\cal L}$ of eq \eqref{eq:sgwb_likelihoodV2} along $\beta$, $I_0$, $n_I$, and we set the variations to zero. 
To simplify next formulas, after taking the variations  we evaluate them   on a frequency band around the reference frequency $f_\star$, hence $\bar I(f=f_\star)\,\simeq\, I_0$. Calling the combination $\hat I_0 =  I_0/(4 \pi f_\star)^2$, we find three conditions, that can be expressed as:
\bea
0&=&\left[ \hat I_0-\frac{(\Gamma^{(1)} \hat {\cal R})}{(\Gamma^{(0)} \Gamma^{(1)})}
\right]+
\nonumber
\\
&&+\frac{\beta\,\hat I_0}{3 (\Gamma^{(0)} \Gamma^{(1)}) }
 \left[(1+n_I) (\Gamma^{(0)} \Gamma^{(0)})+3 (n_I-2) (\Gamma^{(0)} \Gamma^{(2)})+3 (n_I-1)(\Gamma^{(1)} \Gamma^{(1)})  \right]
 \nonumber
\\
&&-\frac{\beta}{3 (\Gamma^{(0)} \Gamma^{(1)}) }
  \left[ (1+n_I) (\Gamma^{(0)} \hat {\cal R}) +3 (n_I-2)(\Gamma^{(2)} \hat {\cal R}) \right]
  \,,
\label{condp1}
\\
0&=&\left[ \hat I_0-\frac{(\Gamma^{(0)} \hat {\cal R})}{(\Gamma^{(0)} \Gamma^{(0)})}
\right]+\frac{\beta\,(n_I-1)}{ (\Gamma^{(0)} \Gamma^{(0)}) }
 \left[2 \hat I_0 (\Gamma^{(0)} \Gamma^{(1)})- (\Gamma^{(1)}\hat {\cal R}) \right]
 \,,
\label{condp2}
\\
0&=&\left[ \hat I_0-\frac{(\Gamma^{(1)} \hat {\cal R})}{(\Gamma^{(0)} \Gamma^{(1)})}
\right]+\nonumber
\\
&&-\frac{\beta\,\hat I_0}{6 (\Gamma^{(0)} \Gamma^{(1)}) }
 \left\{9(\Gamma^{(0)} \Gamma^{(2)})+6  (\Gamma^{(1)} \Gamma^{(1)})
 - 2 n_I \left[
  (\Gamma^{(0)} \Gamma^{(0)})+3  (\Gamma^{(0)} \Gamma^{(2)}) +3
   (\Gamma^{(1)} \Gamma^{(1)})  
  \right]  \right\}
 \nonumber
\\
&&+\frac{\beta}{6 (\Gamma^{(0)} \Gamma^{(1)}) }
  \left[ 3(3-2 n_I) (\Gamma^{(2)} \hat {\cal R}) -2  n_I(\Gamma^{(0)} \hat {\cal R}) \right]
  \,.
\label{condp3}
\eea
In writing the previous expressions, we adopt the abbreviations
\be
(\Gamma^{(0)} \hat {\cal R})\,\equiv\,\sum_A \Gamma^{(0)}_A
\,
 \hat {\cal R}_A
\ee
and analog ones  for the other  round parenthesis. In the previous
 formulas, $ \hat{\cal R}_A$ indicate PTA measurements, while the overlap functions  $\Gamma^{(i)}_A$, for $i=0,1,2$,  depend
 only on the geometrical configuration of the pulsar set. See their definitions in eqs \eqref{defFHD}-\eqref{defF12}.

Taking the difference between \eqref{condp1}  and  \eqref{condp3}, simplifications occur. We
find the following condition involving a linear combination of the GW intensity only
\be
\label{condize}
\hat I_0\,=\,\frac{2 \left( \Gamma^{(0)}  \cdot \hat{\cal R} \right)-
3 \left( \Gamma^{(2)}  \cdot \hat{\cal R} \right)  }{2 \left( \Gamma^{(0)} \cdot  \Gamma^{(0)} \right)-
3  \left( \Gamma^{(2)} \cdot \Gamma^{(0)} \right) }
\,.
\ee
Plugging eq \eqref{condize} in \eqref{condp2}, we find the relation
\bea
&&(\Gamma^{(2)} \Gamma^{(0)}) \,(\Gamma^{(0)} \hat {\cal R})
-(\Gamma^{(2)} \Gamma^{(0)}) \,(\Gamma^{(2)} \hat {\cal R})
\,=
\nonumber
\\
&&= \frac{(1-n_I)\,\beta}{3}\left\{4 (\Gamma^{(1)} \Gamma^{(0)}) \,(\Gamma^{(0)} \hat {\cal R}) 
+\left[ 3  (\Gamma^{(0)} \Gamma^{(2)})- 2 (\Gamma^{(0)} \Gamma^{(0)}) \right] \,(\Gamma^{(1)} \hat {\cal R})-
6  (\Gamma^{(0)} \Gamma^{(1)}) \,(\Gamma^{(2)} \hat {\cal R})
 \right\}\,,
 \nonumber
\\
\label{condibe}
\eea
which provides information on the size of kinematic effects. Suppose to make a set of measurements. We  build the combination in the left hand side of eq \eqref{condibe}. This quantity is expected to be small but generally non zero: it is proportional to the parameter $\beta$ as dictated by the right hand side of this relation. Hence, 
eq \eqref{condibe} suggests
 a practical way to estimate the size of the small parameter $\beta$ by 
 appropriately 
 assembling PTA measurements.  While the formulas discussed so far are evaluated
 at a frequency around the reference frequency $f_\star$, it is straightforward to generalize
 them to arbitrary frequencies, and take the sum over all frequency bins.
 Notice also that if the pulsar positions $\hat x_a$ are orthogonal to the frame velocity $\hat v$, then the functions
$\Gamma^{(1,2)}_A$ vanish. In this case, the previous equality \eqref{condibe} is trivially satisfied, and 
provides no information. 

We now turn to estimating the errors in the measurements using the Fisher matrix approach.

\subsection{Fisher forecasts}
\label{sec_fifo}

\begin{figure}[t!]
    \centering
    \includegraphics[width=0.75\linewidth]{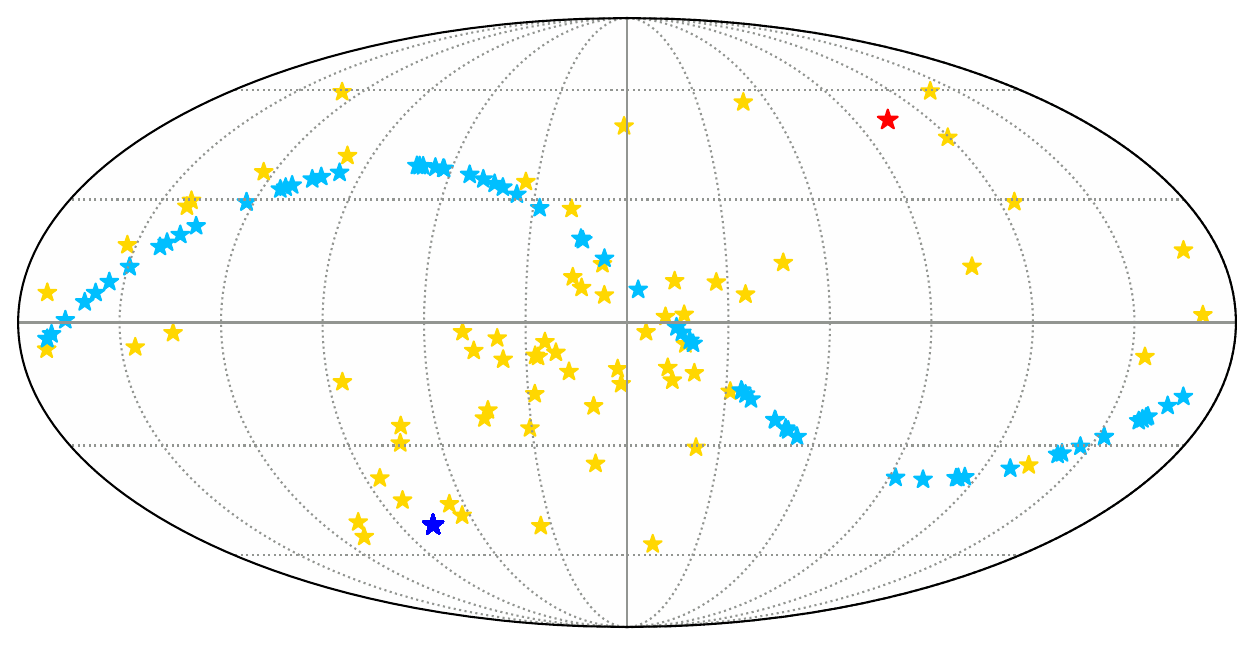}
    \caption{ \it Representation of the sky position of the monitored NANOGrav pulsars (yellow), and the directions (positive and negative) of the velocity vector $\hat v$
    among frames (red and dark blue stars), in galactic coordinates. Light blue stars 
    indicate the position of pulsars in random positions
    orthogonal to the velocity vector.}
    \label{fig:direc}
\end{figure}

The Fisher formalism, based on a manipulation of second derivatives of the log likelihood, allows one to make forecasts of the capabilities of a given experiment (current or future) to measure a given set of parameters. We now
 investigate how an expansion of current pulsar set of NANOGrav collaboration \cite{NANOGrav:2023ctt,NANOGrav:2023hde,the_nanograv_collaboration_2023_8092346,the_nanograv_collaboration_2023_8423265} improves the prospects to detect Doppler anisotropies, if certain conditions are satisfied. See Fig \ref{fig:direc} for a plot
of NANOGrav pulsar positions  in the sky, along as the direction of the velocity vector $\hat v$ as measured
by CMB experiments \cite{Planck:2018vyg}. 

In fact,
currently  the isotropic part of the SGWB is only detected with an 
$\SNR\sim 5$ in the NG15 data set~\cite{NANOGrav:2023gor}. Given that the 
size of dipole anisotropy is at  per mil level relative to the monopole, we do not expect the data to provide accurate information
on kinematic anisotropies at this stage: this is indeed what we found in plot \ref{fig:NG15_upper limts}. However, the Fisher formalism allows us to quantify what gain in sensitivity is required, for the measurement of the kinematic dipole. 

\smallskip

We model our intensity signal as a power-law, as in eq \eqref{defplf}, and we are interested in estimating
the quantities $I_0$ and $\beta$ at first order in a Taylor expansion in $\beta$. Hence, we focus
on the kinematic dipole.\footnote{See appendix \ref{app:A} for the Fisher matrix with the quadrupole and monopole correction included.} The Fisher matrix \eqref{deffim}, when evaluated around the reference frequency $f_\star$ of eq \eqref{defplf}, reads 
\bea
    \fc_{ij}(f_\star) &=& \frac{1}{(4\pi f_\star)^4}\sum_{A}\frac{2T_{A}\Delta f}{\left(R_A^N\right)^2 }
        \,\begin{bmatrix}
        (\beta \kappa\,\Gamma^{(1)}_{A}+\Gamma^{(0)}_{A})^2 & I_0\kappa\,\Gamma^{(1)}_{A}(\beta\kappa\Gamma^{(1)}_{A}+\Gamma^{(0)}_{A}) \\
        I_0 \kappa\,\Gamma^{(1)}_{A}(\beta\kappa\Gamma^{(1)}_{A}+\Gamma^{(0)}_{A}) & (I_0 \kappa\,\Gamma^{(1)}_{A})^2
        \,,
    \end{bmatrix}
    \nonumber\\
    \label{fishrea}
\eea
where the  overlap functions $\Gamma^{(0,1)}_A$ are given in eqs \eqref{defFHD}, \eqref{defF12}, and we introduce 
\be
\label{defkap}
\kappa=n_I-1\,.
\ee 
  Eq \eqref{fishrea} corresponds to the Fisher matrix for a given frequency $f$. The full Fisher matrix is obtained by then summing over the individual frequency bins
\begin{align}
    \fc_{ij} = \sum_f \fc_{ij}(f)\,.
\end{align}

For our fiducial model, we assume~\footnote{These correspond to the the amplitude observed by NANOGrav~\cite{NANOGrav:2023gor} at the chosen reference frequency for a fixed $\gamma=13/3$ and the dipole parameters are the ones measured from the CMB~\cite{Planck:2018vyg}. For plotting
our results we adopt the  conventions of expressing the the intensity magnitude and slope in terms of the parameters $A$ and $\gamma$, as explained in Appendix \ref{app_conv}.}  the following parameter reference values for our plots (see \eqref{I0equivalenceA}): $\log_{10} A = -14.6$ and $\gamma=13/3$.
\bea
I_0 &=& 9.9\times 10^{-23}\,\hskip0.5cm, \hskip0.5cm
\,f_{\star} \,=\, 1/\mathrm{year},
\eea
\bea
\gamma = 13/3 \hskip0.5cm\Longleftrightarrow \hskip0.5cm n_I = -7/3,
\eea
 and 
 \bea
 \beta &=& 1.23\times 10^{-3}
 \hskip0.5cm, \hskip0.5cm
\,
 \hat{v} \,=\, (l,b)=(264^{\circ},48^{\circ})
 \eea
  for the frame velocity amplitude and direction (the latter  in galactic
coordinates).

\smallskip

We can now discuss how
our general approach allows to forecast how  further developments in PTA experiments, for example
monitoring more pulsars, can improve the sensitivity of kinematic anisotropies. As explained
in section \ref{sec_basicf}, the  response of PTA experiments to kinematic anisotropies 
substantially depends on the position of the pulsars with respect to the velocity vector $\hat v$ among frames. We consider two cases including additional pulsars to the existing data set, see Fig \ref{fig:direc}. In the first case, we add 67 extra pulsars beside the 67 ones of NANOGrav collaboration, 
for simplicity each characterized by the same noise as an NG15 pulsar. The additional pulsars are all located in   directions orthogonal  to $\hat v$.  Doubling the  number of  pulsars  allows us to build many more pulsar pairs.

\begin{figure}[t!]
    \centering
    \includegraphics[width=0.4\linewidth]{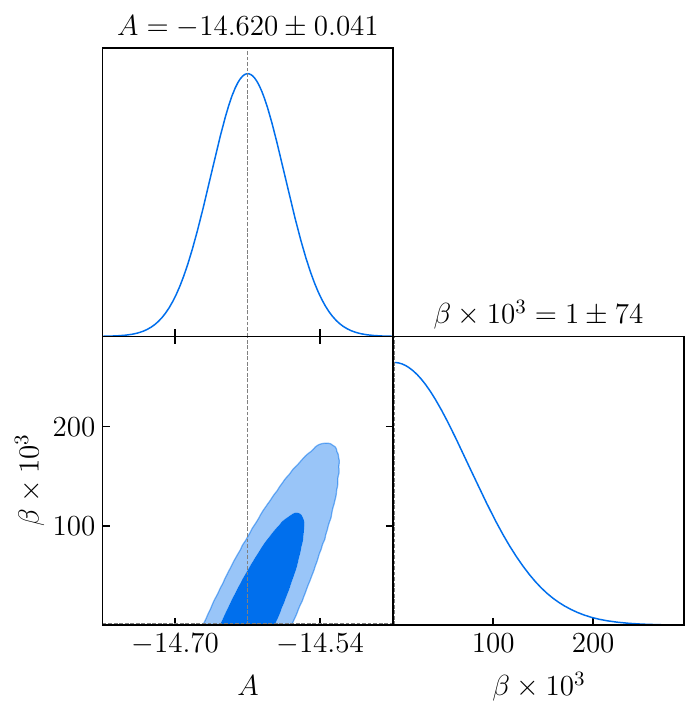}
        \includegraphics[width=0.4\linewidth]{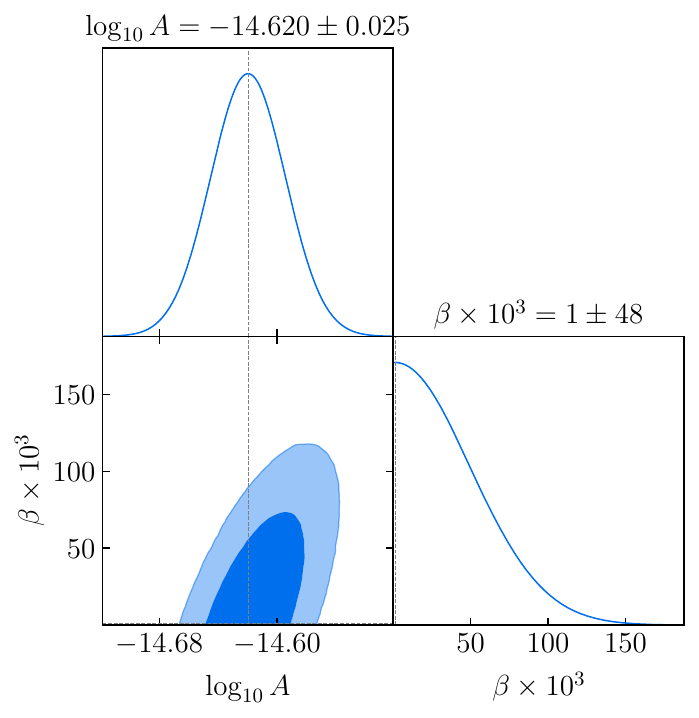}
    \caption{ \it Results of our analysis when adding 67 additional pulsars to the NG15 data set, in directions
    orthogonal ({\bf left panel}) and parallel ({\bf right panel}) to the velocity vector $\hat v$, as explained in the main text. 
    }
    \label{fig:adNG15_fisher}
\end{figure}

However, when pairing among the additional pulsars only, one finds that the sensitivity to Doppler anisotropies vanish, because the contributions
$\Gamma^{(1,2)}_{ab}=0$ in eqs \eqref{defF11}-\eqref{defF12}  since the pulsars are orthogonal to $\hat v$. As a consequence, although we somewhat gain in sensitivity, the error in the parameter $\beta$ is still large. See Fig \ref{fig:adNG15_fisher}, left panel. 
We then consider a second case, where  we add 67 extra pulsars beside the 67 ones of NANOGrav collaboration, 
 all located in a  direction {\it parallel}  to $\hat v$. We represent the result in Fig \ref{fig:adNG15_fisher}, right panel. 
 In this case, formulas  \eqref{defF11}-\eqref{defF12}  indicate that the sensitivity to kinematic
 anisotropies should increase. In fact, we find that the error bars considerably reduce with respect to the first case discussed
 previously. Nevertheless, for both cases,  working with order ${\cal O}(100)$ pulsars do not seem sufficient to achieve the desired per mil sensitivity on the value of $\beta$. In what comes next, we discuss futuristic PTA experiments that might be able to do so, by monitoring many more pulsars.

\section{Doppler anisotropies and future PTA data}
\label{sec:Fisher_ideal}

After estimating the prospects for the present generation
of  PTA experiments to detect
kinematic anisotropies in the SGWB, we now look towards the future,
 and we consider
futuristic PTA experiments monitoring a very large number of pulsars. We assume to
observe   $N_{\rm psr}\gg{\cal O}(100)$ identical pulsars distributed isotropically across the sky, and observed  for the same time period $T$. This is an optimistic condition, but in some approximation
it might be  achieved in the SKA era \cite{Janssen:2014dka}. 
Given that the pulsars are assumed to be isotropically distributed, in this section we do not exploit
their particular positions in the sky. Instead we investigate how a large number
of pulsars improve the prospect to detect Doppler effects and consequently the magnitude and direction of the kinematic dipole. 
 In first approximation, we can expect that
 the signal-to-noise ratio scales as the square
 root of the time of observation, times
 the square root of the number $N_{\rm pair}$ of pulsar pairs \cite{Anholm:2008wy}. Since the latter
 scale as \mbox{$N_{\rm pair} \,=\,N_{\rm psr}(N_{\rm psr}-1)/2 \sim N_{\rm psr}^2/2$} for a large number of pulsars, increasing the quantity of pulsars considerably improves the prospects of detection.

For 
our forecasts we use the same approach of section \ref{sec_info}.  Additionally, in order to handle our formulas more
easily, we use the techniques developed in \cite{Ali-Haimoud:2020ozu}. We express the GW-induced
two point correlators among time residuals as in eq \eqref{r2pt1}: i.e., we do not integrate
over directions, and we make use of the tensor $\gamma_{ab}$ of eq \eqref{contra}. The Fisher
matrix, evaluated at frequency $f$, reads
\begin{align}
    \fc_{ij}(f) = \frac{2T\Delta f}{(4\pi f \sigma)^4}N_{\rm pair}\times \frac{1}{N_{\rm pair}}\sum_{A} \gamma_{A} \cdot \frac{\partial I}{\partial \Theta_i} \gamma_{A} \cdot \frac{\partial I}{\partial \Theta_j}\,, 
    \label{fishex2}
    \,
\end{align}
where $\sigma^2$ is the (common) noise variance.
{ We assume that all pulsars have
the same intrinsic noise in order to   carry on analytical arguments in this section.}
In writing \eqref{fishex2}, we vary over the parameters
we wish to determine, as explained in section \ref{sec_info}. 

Using the results of \cite{Ali-Haimoud:2020ozu},
 in the limit of large $N_{\rm psr}$ pulsars distributed isotropically across the sky, we
 can write 
 \begin{align}
    \frac{1}{N_{\rm pair}} \sum_{A} \gamma_{A} \cdot \frac{\partial I}{\partial \Theta_i} \gamma_{A}\cdot \frac{\partial I}{\partial \Theta_j} = \int \frac{d^2 a}{4\pi} \frac{d^2 b}{4\pi} \int \frac{d^2 n}{4\pi} \frac{d^2 n'}{4\pi} \gamma_{ab}(\hn)\gamma_{ab}(\hn') \frac{\partial I(\hn)}{\partial \Theta_i}\frac{\partial I(\hn')}{\partial \Theta_j}\,,
\end{align}
where we  replace the sum over pulsar pairs with an integral over the pulsar pair directions. Then, exchanging the order of the integrals,  we can define a new function $ \mathcal{F}_{\infty}$ as 
    \begin{align}
      \lim_{N_{\rm psr}\to \infty}  \frac{1}{N_{\rm pair}} \sum_{ab} \gamma_{ab}(\hn) \,\gamma_{ab}(\hn') &= \int \frac{d^2 a}{4\pi} \frac{d^2 b}{4\pi} \gamma_{ab}(\hn) \,\gamma_{ab}(\hn') 
   \,   \equiv\, \mathcal{F}_{\infty}(\hn\cdot\hn')
   \,.
    \end{align}
Defining $\chi \equiv \hn \cdot \hn'$, the function  $\mathcal{F}_{\infty}(\chi)$
    results in~\cite{Ali-Haimoud:2020ozu}
  \begin{align}
    \mathcal{F}_{\infty}(\chi) = \frac{16}{9 (1 + \chi)^2} 
\times &\Bigg{[} \left(\frac{1- \chi^2}{4} +  2 - \chi + 3\frac{1 - \chi}{1 + \chi} \log\frac{1-\chi}{2} \right)^2  \nonumber\\
&+ \left(2-\chi  + 3\frac{1 - \chi}{1 + \chi} \log\frac{1-\chi}{2}\right)^2\Bigg{]}\,. \label{eq:Fdense}
\end{align}
We emphasise that $\chi$ controls the angle between two GW directions. It  is distinct from
the angle $\zeta_{ab}$ used in eq \eqref{defyab}. 

At this point, given the isotropic distribution of the system, it is convenient to expand the 
quantities involved in spherical harmonics, and use their orthogonality properties. For the function $  \mathcal{F}_{\infty}(\hn\cdot\hn') $ we have
\begin{align}
 \label{eq:sph_IF}
    \mathcal{F}_{\infty}(\hn \cdot \hn') &= 4\pi \sum_{\ell m} \fc_{\ell}\,Y_{\ell m}(\hn)Y_{\ell m}^*(\hn')
\end{align}
while for the anisotropic intensity -- whose anisotropy is induced by kinematic effects as 
in eq \eqref{tayexa} -- we can write
\begin{align}
    \label{eq:sph_IF2}
    I(f,\hn)&=  \bar{I}(f) \,\left[1+\frac{4\pi}{3}\beta (n_I-1)\sum_{m}Y_{1m}(\hn)Y_{1m}^* (\hat v)\right]
\,.
\end{align}
 We consider only dipolar anisotropies up to order $\beta$ (but notice that in the previous
fornula $n_I$ can depend on frequency). The first three  coefficients of $\fc_{\infty}$ in eq \eqref{eq:sph_IF} read $\fc_0 = 4/27,\, \fc_1 = 32 \zeta (3)-\frac{346}{9} \approx 0.02,\, \fc_2 = \frac{46742}{135}-288 \zeta (3) \approx 0.04$, with $\zeta(n)$  the Riemann zeta function.

Substituting expression \eqref{eq:sph_IF2} into our formula \eqref{fishex2} for the Fisher
matrix  evaluated
at a given frequency $f$, we find
\begin{align}
    \fc_{ij}(f) = \frac{2T\Delta f }{(4\pi f \sigma)^4}N_{\rm pair}
    \times \left[ \fc_0\frac{\partial \bar I}{\partial \Theta_i}\frac{\partial \bar I}{\partial \Theta_j} + \frac{4\pi\fc_1}{9}\sum_{m}\frac{\partial I_{1m}}{\partial \Theta_i}\frac{\partial I_{1m}^{*}}{\partial \Theta_j}\right]
    \,,
\end{align}
where
\begin{align}
    I_{1m} \equiv  \bar I \,\beta\,\kappa \,Y_{1m}(\hv{v})
    \,,
\end{align}
(recall our definition of the parameter $\kappa=n_I-1$ in eq \eqref{defkap}). 

We vary along the  intensity amplitude $\bar I$ -- for simplicity we then evaluate
the Fisher matrix 
 at a reference frequency $f_\star$ --, along with the
size of the frame velocity $\beta$, as well as along
 the dipole direction $\{\theta,\phi\}$ (which we later translate to galactic coordinates). We find
\begin{align}
    \label{eq:Fisher_4param}
    \fc_{ij}(f_\star) =&  \frac{2T\Delta f }{(4\pi f \sigma)^4}N_{\rm pair}\times
    \begin{bmatrix}
        \fc_0 + \frac{\fc_1 \kappa^2\beta^2}{3}  &  \frac{\bar I\beta \fc_1\kappa^2 }{3}& 0 & 0\\
        \frac{\bar I\beta \fc_1\kappa^2}{3}  & \frac{\bar I^2 {\fc_1}\kappa^2}{3}   & 0 & 0\\
        0 & 0 & \frac{\fc_1 \bar I^2 \kappa^2 \beta^2}{3} & 0 \\
        0 & 0 & 0 & \frac{\fc_1 \bar I^2 \kappa^2 \beta^2\sin^2 \theta}{3}
    \end{bmatrix} 
    \,.
\end{align}
Once again, the full Fisher matrix will then be obtained by summing over the individual frequency bins.

\subsection{ Estimating the parameters in specific examples}

We now apply the previous formulas to specific examples, in order 
to forecast in the idealized
case of this section the accuracy in determining the  model parameters, including the 
frequency dependence of the background.

\subsubsection{  Power-law scenario}
\label{subsubpl}

Let us start with the example of a power-law frequency dependence of
the SGWB in its rest frame, as the one of eq \eqref{defplf}. Kinematic anisotropies, up to order $\beta$, lead to an anisotropic intensity of the form
\begin{align}
\label{eq:plaw}
I(f,\hn)=I_0(f/f_\star)^{n_I}(1+\beta\,\kappa\,\hn\cdot\hv{v})
\,,
\end{align}
with $\kappa=n_I-1$ as in eq \eqref{defkap}, and $f_\star$ a reference frequency. 
Adding up the Fisher matrix over the frequency bins leads to 
\begin{align}
    \fc_{ij} =     \sum_{f}\frac{2T\Delta f }{(4\pi f \sigma)^4}N_{\rm pair} \left(\frac{f}{f_{\star} }\right)^{2n_I}    \times
    \begin{bmatrix}
        \fc_0 + \frac{\fc_1 \kappa^2\beta^2 }{3} &  \frac{I_0\beta \kappa^2\fc_1}{3} & 0 & 0\\
        \frac{I_0\beta \kappa^2\fc_1 }{3} & \frac{I_0^2 \kappa^2{\fc_1}}{3}   & 0 & 0\\
        0 & 0 & \frac{\fc_1 I_0^2\kappa^2 \beta^2 }{3}& 0 \\
        0 & 0 & 0 & \frac{\fc_1\,I_0^2\kappa^2 \beta^2\sin^2 \theta}{3}
    \end{bmatrix}\,.
\end{align}

We now
focus on the component ${\cal F}_{\beta\beta}$, in order to Fisher estimate the
error on the velocity amplitude (see section \ref{sec_fifo}). 
We obtain
\begin{align}
    \fc_{\beta\beta} = \sum_{f}\frac{2T\Delta f }{(4\pi f \sigma)^4}N_{\rm pair}\left(\frac{f}{f_{\star} }\right)^{2-2\kappa} \frac{I_0^2 \kappa^2 \fc_1}{3}= \SNR_{\rm iso,\rm tot}^2 \frac{\kappa^2\fc_1}{3\fc_0}
    \,,
\end{align}
where we identify the signal-to-noise ratio of the isotropic part as
\begin{align}
    \SNR_{\rm iso,tot}^2 =  \sum_{f}\frac{2T\Delta f }{(4\pi f \sigma)^4}N_{\rm pair} \left(\frac{f}{f_{\rm ref} }\right)^{2-2\kappa} I_0^2 \fc_0\, 
        \,.
\end{align}
Therefore, the error in the measurement of $\beta$ becomes (at lowest order in $\beta$), 
\begin{align}
    \label{eq:beta_err_snr}
    \Delta \beta = [F^{-1}]^{1/2}_{\beta\beta} \approx \left[\frac{3\fc_0}{\SNR_{\rm iso,\rm tot}^2 \kappa^2 \fc_1 }\right]^{1/2}\,.
\end{align}
{Such estimates for the minimum detectable magnitude of the dipole anisotropy (not specific to the kinematic dipole) have been previously derived in~\cite{Hotinli:2019tpc,Ali-Haimoud:2020ozu}. Note however that those estimates are presented in terms of the overall dipole magnitude, which in our case corresponds to the product $\kappa \beta$. Thus, estimates of  $\beta$ depend on the spectral slope parameter $\kappa\,=\,n_I-1$.}
As we know, 
CMB measurements give a value $\beta = 1.2\times 10^{-3}$, which implies that for PTA experiments to measure this parameter with a precision of $\Delta\beta \sim 10^{-3}$ (in order for being competitive with other data sets)  we must ensure  at least
 \begin{align}
    \SNR_{\rm iso,tot} \sim {10^4}\,. 
\end{align}
Starting from these considerations, we can derive quantitative estimates for the error bars
on the parameters involved, making some assumptions on the PTA measurements.

\begin{figure}
    \centering
    \includegraphics[width=0.35\linewidth,valign=m]{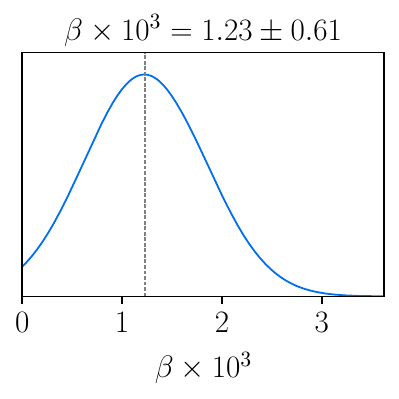}
     \includegraphics[width=0.45\linewidth,valign=m]{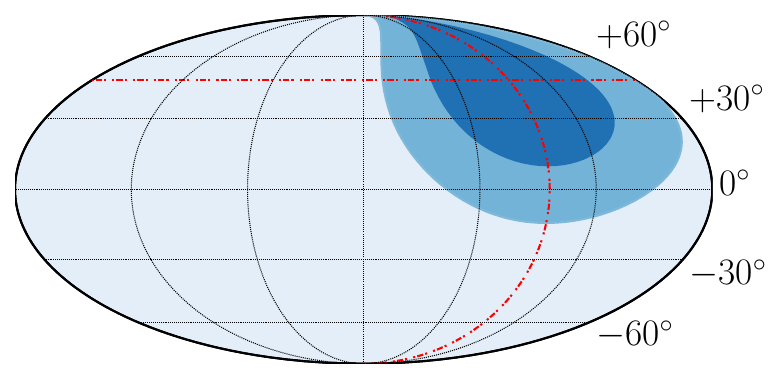}
    \caption{\it Fisher forecast for 
    $\beta$
and the dipole direction parameters, as discussed in the main text around eq \eqref{resila1}.}
    \label{fig:Fisher_test}
\end{figure}

We take the time of observation $T_{\rm obs}=20\,\mathrm{years}$ and for the pulsar white noise parameters, we fix \mbox{$T_{\rm cad} = \mathrm{year}/20$} as the cadence of the timing observations, $\Delta t_{\rm rms} =100\,\mathrm{ns}$ the rms error of the timing residuals~\cite{Hazboun:2019vhv}. For the red-noise parameters we take $A_{\rm RN}= {2\times 10^{-15}}$ and $\alpha_{\rm RN}=-2/3$.\footnote{{This choice of red-noise parameters ensures that, to good approximation, we remain in the weak-signal limit, where the results of eq.~\eqref{eq:cov_weaksignal} and onwards apply.   This corresponds
to a red-noise amplitude somewhat towards the higher end of what has been observed in
the NANOGrav pulsar set \cite{NANOGrav:2023hde}. We also discuss below what we expect would happen for a lower  noise level. }}  We then use \texttt{Hasasia}~\cite{Hazboun2019Hasasia} to generate the noise curve. We perform a frequency binning, taking $\Delta f = 1/T_{\rm obs}$ and $f\in [1/T_{\rm obs}, 20/T_{\rm obs}]$. For our fiducial parameters, we assume the same values as in section \ref{sec_fifo}. Fixing this noise level, we choose the number of pulsars so as to obtain {at least a 95\% confidence level detection} 
of $\beta$. We find that for this we need {4000} identical pulsars with the above noise properties to obtain such a precision. For this configuration, we plot the joint $68\%$ and $95\%$ confidence intervals for $\log_{10} A,\,\beta$ 
and the dipole direction parameters in figure \ref{fig:Fisher_test}. The marginalised parameter limits we obtain at $68\%$ level are
{
\begin{align}
\label{resila1}
    \log_{10} A &= -14.6\pm 0.000094\,, \\
    \beta &= (1.23\pm 0.61)\times 10^{-3}\,,\\
    b &= 48\pm 30\,,\\
    l &= 264\pm 40\,.
\end{align}}
\noindent
where the two parameters $(l,b)$ indicate the dipole direction in galactic coordinates.
Hence, we conclude that two decades of observations as well as the observation of {4000} pulsars in this idealized case allows us to measure
parameters controlling 
Doppler effects, with sufficient accuracy so that the results can be compared to other (CMB and LSS) data sets. In fact, since $N_{\rm pair}$ scales quadratically with $N_{\rm psrs}$, the accuracy in the determination of the dipole parameters scales linearly with the number of pulsars in this idealised limit. {
The  number $N_{\rm psrs}$  of pulsars we need to monitor 
 is large
even for future experiments
in  SKA \cite{Janssen:2014dka}, hence our scenario is futuristic. Nevertheless,
such large numbers be achievable with astrometry (see e.g. \cite{Book:2010pf}).   } 

{
 The measurement error on $\beta$ scales as $\Delta \beta \propto 1/\SNR$ with the following $\SNR$ scaling in the weak-signal limit~\cite{Siemens:2013zla}
\begin{align}
    \SNR \propto \frac{I N_{\rm psr} T^{2-n_I}  }{\sigma^2}\,,
\end{align} 
which suggests that decreases in the noise level can significantly improve the prospects of detection of the kinematic dipole. However, this is only true in the weak-signal limit. Given that several pulsars monitored by current PTA experiments  are already in the intermediate signal regime, it is likely that future PTA observations 
will reach
 the strong-signal regime where $\SNR \propto N_{\rm psr}\sqrt{T}$~\cite{Siemens:2013zla}. This means that decreases in the intrinsic pulsar noise level will only provide minor, rather than significant improvements to the constraints on the dipole magnitude.
 For example, going from \mbox{$\sigma^2 \to \sigma^2/10$} will not imply \mbox{$\Delta\beta \to \Delta \beta /10$}; rather,  a more modest improvement will be expected.  

Irrespective of the regime in which future PTAs will operate, increasing the number of monitored pulsars  will still be the best strategy to improve  constraints on the dipole anisotropy, and on the size of anisotropies in general. The position of the pulsars being observed will also be crucial and as we have seen from section~\ref{sec_fifo}, a PTA with pulsars in the dipole direction will be much more suited to detecting the dipole as compared to pulsars distributed uniformly. Further improvements could be also be brought about by increasing the frequency of observations (cadence), thereby allowing for the use of information from the high frequency part of the spectrum as well.} 

\smallskip

We also explored whether in this particular example we can use kinematic anisotropies to better constrain the slope of the spectrum $\kappa$, with respect to monopole measurements only. The answer is negative since for power law case information on the dipole anisotropy  does not break any degeneracy between amplitude and slope of the spectrum. Instead, kinematic anisotropies are more informative with respect to the spectra slope for backgrounds more complex than a power law, as we are going to examine next.

\subsubsection{ Log-normal frequency spectrum}
\label{sec_logn}

Our formalism is sufficiently flexible to allow us to
 go beyond the case of a power-law. 
\begin{figure}[t!]
    \centering
    \includegraphics[width=0.35\linewidth]{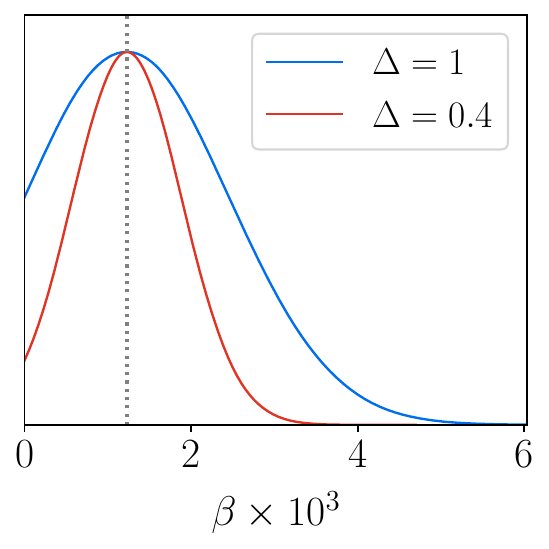}
    \caption{ \it Fisher forecast of the dipole magnitude $\beta$ for log-normal spectrum of section \ref{sec_logn}. The parameter choices are shown in table~\ref{tab:lgn_params}, we obtain $\beta = (1.23\pm {1.5})\times 10^{-3}$ for $\Delta=1$ and 
$\beta = (1.23\pm {0.4})\times 10^{-3}$ for $\Delta=0.4$ (at $68\%$ C.L). }
    \label{fig:Fisher_lgn}
\end{figure}
Let us consider a log-normal
 frequency profile in the SGWB rest frame, with 
a peaked frequency feature:
 \begin{align}
    \bar I(f) = \frac{I_0}{\sqrt{2\pi}\Delta}
    \exp\left(-\frac{\ln^2(f/f_0)}{2\Delta^2}\right)\,,
\end{align}
with $\Delta \ll 1$ corresponding to a sharply peaked narrow spectrum and $\Delta \gtrsim 1$ to a relatively broader spectrum.

The resulting Fisher forecasts, obtained
following the same procedure of section \ref{subsubpl}, are plotted in figure \ref{fig:Fisher_lgn},
for the two cases $\Delta=0.4$ and $\Delta=1$.

To illustrate our point regarding the role of the frequency dependence, we focus only on the estimation of the dipole magnitude $\beta$.\footnote{The results can be easily extended to include the dipole direction parameters using equation \eqref{eq:Fisher_4param}.}
We consider two cases, namely a narrow and a broad spectrum with the parameter choices for both presented in table~\ref{tab:lgn_params}.

\begin{table}[h]
	\centering
	\renewcommand\arraystretch{1.1}
	\begin{tabular}{|c|c|c|}
		\hline
		\textbf{Parameter} & \textbf{Narrow} & \textbf{Broad} \\
		\hline
		\boldmath{$\log_{10}I_0$}   &  {$-20.4$} &{-20.4}\\
		\boldmath{$f_0$} & $0.2/\mathrm{year}$ & $0.2/\mathrm{year}$\\
		\boldmath{$\Delta$} & $0.4$ & $1$\\
		\hline
	\end{tabular}
	\caption{Parameter values for the narrow and broad log-normal power spectra.  }
	\label{tab:lgn_params}
\end{table}

The parameters are chosen in a manner 
such that the isotropic SNR for both cases is the same. We learn that sharper features in the frequency spectrum enhance  kinematic effects, making them easier to detect in the $\Delta=0.4$ case compared to the $\Delta = 1$ case. 
This is a particular example among many showing how the possibly rich frequency dependence  of the GW spectrum can affect the detection of Doppler anisotropies.

\section{Conclusions}
\label{sec:conclusion}

 The recent detection of a stochastic gravitational wave background by PTA collaborations represents a significant milestone in the field of GW astronomy. Given that the origin of this background is currently unknown, it is a task for  theorists to identify signatures that could be used to discriminate between astrophysical and cosmological origin of the signal. 
 In the near future, the IPTA DR3 release is likely to lead to more precise measurements of the amplitude and spectral shape and may provide hints to origin of the SGWB~\cite{InternationalPulsarTimingArray:2023mzf}. Another key signature is the anisotropy of the SGWB intensity, with astrophysical models predicting a much larger level of anisotropy compared to cosmological ones. Thus, future observation of anisotropies may decisively tilt the scales in favour of a cosmological or astrophysical original of the PTA signal. 

Promisingly, recent forecasts of the detectability of astrophysical anisotropies suggest that a significant fraction of the predicted level of anisotropy may be detected in the current or near future data~\cite{NANOGrav:2023tcn}. Although the prospects for detecting the intrinsic anisotropy of cosmological backgrounds are currently quite dire, 
the kinematic dipole anisotropy which is generated due to our motion with respect to  the SGWB rest frame {is expected to be much larger than the intrinsic anisotropy for such backgrounds}. Using the NG15 dataset, we are able to place an upper limit {$\beta<0.297$ at $95\%$ C.L.} on the magnitude of this dipole, assuming a cosmological origin for the signal and the dipole direction to be the same as inferred from the CMB. We also 
{quantify the number of pulsars that will be required with future datasets to detect the kinematic dipole, finding that it will require $N \approx 4000$ pulsars, for the pulsar red noise parameters that we have chosen.} 
{We stress that our results  are derived under the weak-signal approximation, which may not hold for futuristic PTAs. Thus, it would be interesting to explore how the detection forecasts for the dipole anisotropy change in the opposite regime where the signal might dominate,  even though the results are not expected to be notably different. Also, it would be important to exploit the dependence on the position of pulsars (as discussed in sections \ref{sec_basicf} and \ref{sec_pres}) to improve the forecasts and reduce error bars in  PTA systems, going beyond some of the analytical approximations
used in section \ref{sec:Fisher_ideal}. Moreover,  it would be interesting to extend and apply some of our techniques developed for PTA to the case of astrometry, where a large number of stars can be monitored with the aim of  detecting GW.
} 

{Future observations of SGWB with PTAs may also allow us to test} whether the rest frames of the CMB and cosmological SGWB coincide. While one would naturally expect this to be the case, the emergence of the recent CMB-LSS dipole tension underscores the importance of verifying this assumption experimentally. In this way, observation of the the SGWB kinematic dipole could enable us to not only pinpoint the origin of the PTA signal, but also serve as an independent test of the cosmological principle. Last, but not least, Doppler anisotropies can also provide independent tests of modifications of Einstein gravity. The   search for their presence with PTA experiments is definitely worth pursuing.

\subsection*{Acknowledgments}

We are partially funded by the STFC grants ST/T000813/1 and ST/X000648/1. 
We also acknowledge the support of the Supercomputing Wales project, which is part-funded by the European Regional Development Fund (ERDF) via Welsh Government. For the purpose of open access, the authors have applied a Creative Commons Attribution licence to any Author Accepted Manuscript version arising.

\begin{appendix}
\section{ Conventions on the SGWB parameters}
\label{app_conv}
The spectral energy density parameter $\Omegagw$ is related to the intensity $I$ through
    \begin{align}
        \Omegagw  = \frac{4\pi^2 f^3}{3H_0^2}I \,.
    \end{align}
    Besides the GW energy density,
another important quantity is the characteristic strain $h_c$, which we parametrise as a power law 
\begin{align}\label{AIdef}
    h_c(f) \equiv \sqrt{2 f I(f)} = A_{\rm }\left(\frac{f}{f_{\rm ref}}\right)^\alpha\,.
\end{align}
Note that for a background of supermassive black hole binaries, the expected parameters for the characteristic strain are $\alpha=-2/3$ and $A_{\rm }\approx 10^{-15}$. PTA collaborations typically report their results in terms of the parameters $A_{\rm },\gamma$ where 
\be\label{gammadef}
\gamma\equiv 3-2\alpha.
\ee
The NG15 results for the amplitude $A_{\rm }$, when fixing $\alpha=-2/3$ and taking a reference frequency of $f_{\rm ref}=1/\mathrm{year}$ are~\cite{NANOGrav:2023gor}
\begin{align}
    \label{eq:A_NG15}
    A_{\rm } = 2.4^{+0.7}_{-0.6}\times 10^{-15}\, (\text{at } 90\% \text{ C.L.})\,. 
\end{align}
For such a power law characteristic strain model, the spectral energy density can be expressed as
\begin{align}
       \Omegagw(f) = \frac{2\pi^2 f^2}{3H_0^2} A_{\rm }^2 \left(\frac{f}{f_{\rm ref}}\right)^{2\alpha} = \Omega_{\mathrm{GW},0} \left(\frac{f}{f_{\rm ref}}\right)^{n_\Omega}\,,
\end{align}
where
\begin{align}
         \Omega_{\mathrm{GW},0} \equiv  \frac{2\pi^2 }{3H_0^2}f_{\rm ref}^2 A_{\rm }^2,\quad n_{\Omega} \equiv 2\alpha+2\,.
\end{align}
Similarly, for the intensity we have
\begin{align}
    I (f) = \frac{A_{\rm }^2}{2f}\left(\frac{f}{f_{\rm ref}}\right)^{2\alpha} = I_0 \left(\frac{f}{f_{\rm ref}}\right)^{n_I}
    \,,
\end{align}
where
\begin{align}\label{I0equivalenceA}
    I_0\equiv \frac{A_{\rm }^2}{2f_{\rm ref}},\quad n_I \equiv 2\alpha - 1\,.
\end{align}
Plugging in the value of $A_{\rm }$ measured by NANOGrav \eqref{eq:A_NG15}, we have
\begin{align}
    \Omega_{\mathrm{GW},0} \approx 8.1\times 10^{-9},\quad I_0 \approx 
 9.1\times 10^{-23} [s]\,.
\end{align}
If instead, we take as our reference frequency $f_{\rm ref} = 0.1/\mathrm{year}$ we obtain,
\begin{align}
    \Omega_{\mathrm{GW},0} \approx 5.6\times 10^{-10},\quad I_0 \approx 
 6.3\times 10^{-21} [s]\,.
 \end{align}

\section{Covariance matrix in the weak signal limit}\label{appcov} 

We now provide arguments to explain why  the covariance matrix in our likelihood is effectively diagonal in the weak signal limit --  see the discussion after eq \eqref{eq:cov_def}.  Our arguments are not specific to PTA experiments, but they are  also applicable to ground based interferometers. Let $d_I$ denote the data recorded by a given detector\footnote{This can be either the single detector ouput or the cross-correlated output in which case the subscript $I$ denotes either a pulsar or interferometer pair.} and $N_I$ denote its noise PSD. The data is the sum of the signal and noise, i.e. 
\begin{align}
	d_ I = s_I + n_I,\,\quad \qev{d_I} = \qev{s_I},\quad \qev{n_I} = 0\,.
\end{align}
We write the likelihood as
\begin{align}
	-2\ln \lc = \sum_{IJ} (\hat{d}_I - s_I)\,C^{-1}_{IJ}\,(\hat{d}_J- s_J)^T\,,
\end{align}
where the covariance $C_{IJ}$ is given by 
\begin{align}
	C_{IJ} = \qev{d_I d_J} - \qev{d_I}\qev{d_J}\,.
\end{align}
Suppose for simplicity we have only two detectors, so that $C_{IJ}$ is a $2\times 2$ matrix. Then assuming the noise is uncorrelated across detectors, and that the detectors have the same noise variance,  we find 
\begin{align}	
	C_{IJ} = 
	\begin{pmatrix}
	N + \qev{s_I^2} &\qev{s_I s_J} \\
		  \qev{s_I s_J}  & N + \qev{s_J^2}
	\end{pmatrix}\,.
\end{align}
For ease of notation, we assume $\qev{s_I} = \qev{s_J} = s$ and $ \qev{s_I^2}=\qev{s_J^2} = \qev{s_I s_J} = \qev{s^2}$. Evaluating the full likelihood for the signal with the above covariance matrix gives
\begin{align}
	-2\ln \lc = \frac{2(\hat d - s)^2}{N+2s^2}\approx \frac{2(\hat d - s)^2}{N}
\end{align}
in the weak signal limit $s^2 \ll N$. Thus, in this limit we approximate the covariance matrix as effectively diagonal,
\begin{align}
C_{IJ} \approx 
\begin{pmatrix}
	N  & 0\\
	0  & N 
\end{pmatrix}
\,,
\end{align}
as claimed. 

\section{Full Fisher matrix with  contributions up to  \texorpdfstring{order $\beta^2$}{second order in beta}}
\label{app:A}
For convenience,
in this appendix we express   the intensity as
\begin{align}
    I(f,\hn) = I_0 H(f)\left[1+m_I(f)\beta^2 +\xi d_I(f)\beta + \frac{1}{2}(3\xi^2 - 1)q_I(f)\beta^2\right]
    \label{eqforia}
    \,,
\end{align}
with $\xi=\hn\cdot \hv{v}$ and 
\begin{align}
    m_I(f) &= \frac{1}{6}(-1 +n_I^2 + \kappa_I)\,,\\
    d_I(f) &= (1-n_I)\,,\\
    q_I(f) &= \frac{1}{3}[\kappa_I + (n_I-2)(n_I-1)]\,.
\end{align}
We consider two cases. A realistic case with finite number of pulsars, as
in section \ref{sec_pres}, and an idealised case  with a very  large number of pulsars
distributed isotropically in the sky, as in section \ref{sec:Fisher_ideal}. 

\subsection{Realistic case}
The full likelihood with quadrupole and monopole correction added is
\begin{align}
    -2\ln \lc = \mathrm{const.}+\sum_f \sum_{pq} \left(\hat{\rc}_{pq} - \frac{I_0 H(f)(\Gamma_{pq}^{(0)}(1+\beta^2 m_I)+\beta(1-n_I)\Gamma_{pq}^{(1)} + \beta^2 q_I\Gamma^{(2)}_
    {pq})}{(4\pi f)^2} \right)^2\, N^{-1}_{pq}\,,
\end{align}
with 
\begin{align}
    \Gamma^{(2)}_{pq} &\equiv \frac{1}{2}\int \frac{d^2\hn}{4\pi}\, {\gamma_{pq}(\hn)}(3\xi^2 - 1)\,.
\end{align}
We obtain  the following Fisher matrix for $\{I_0,\beta\}$,
\begin{align}
    \fc_{ij,f} = \frac{1}{(4\pi f)^4}\sum_{pq}\frac{2T_{pq}\Delta f}{\sigma_p^2 \sigma_q^2}H^2(f)\,
    \begin{bmatrix}
a_{I_0 I_0} & a_{I_0 \beta} \\
        a_{\beta I_0} & a_{\beta \beta}
        \,,
    \end{bmatrix}
\end{align}
with
\begin{align}
    a_{I_0 I_0} =& \left(  \beta(  \kappa\Gamma^{(1)}_{pq}+q\beta\Gamma^{(2)}_{pq})+\Gamma^{(0)}_{pq} \left(\beta ^2 m+1\right)\right)^2 \approx (\Gamma^{(0)}_{pq})^2 + \oc(\beta) \\
    a_{I_0 \beta} =& \,a_{\beta I_0} = I_0\left(\beta(  \kappa\Gamma^{(1)}_{pq}+q\beta\Gamma^{(2)}_{pq})+\Gamma^{(0)}_{pq} \left(\beta ^2 m+1\right)\right) \left(\kappa  \Gamma^{(1)}_{pq}+2 \beta  \Gamma _0 m+2 \beta  \Gamma^{(2)}_{pq} q\right) \\ 
    \approx & \,I_0\kappa\Gamma^{(0)}\Gamma^{(1)}_{pq}+\oc(\beta) \\
    a_{\beta\beta} =& I_0^2\left(\kappa\Gamma^{(1)}_{pq}+\beta(m\Gamma^{(0)}_{pq}+q\Gamma^{(2)}_{pq})\right)^2 \approx (I_0\kappa \Gamma^{(1)}_{pq})^2 + \oc(\beta)
    \,.
\end{align}
Thus, in the limit $\beta \ll 1$, the contributions of the quadrupole and the monopole correction are subdominant compared to the dipole. However, in the realistic case, the Fisher matrix also depends on the exact configuration of pulsars. Although both the dipole and quadrupole response vanish for pulsars orthogonal to the velocity, the exact dependence is slightly different and it is indeed possible that for certain configurations of pulsars the quadrupole response is as or more important than
the dipole.

\subsection{Idealised case}
Proceeding similarly to section \ref{sec:Fisher_ideal} we obtain
\begin{align}
    \fc_{ij,f} &= \frac{2T\Delta f }{(4\pi f \sigma)^4}N_{\rm pair}H^2(f)  \times
    \begin{bmatrix}
        a_{I_0 I_0} &  a_{I_0\beta} & 0 & 0\\
        a_{\beta I_0}  & a_{\beta \beta} & 0 & 0\\
        0 & 0 & a_{\theta\theta} & 0 \\
        0 & 0 & 0 & a_{\phi\phi}
    \end{bmatrix}      
    \,,
\end{align}
where
\begin{align}
    a_{I_0 I_0} &= \fc_0 \left(m_I \beta^2+1\right)^2+\frac{1}{3} \beta ^2 \fc_1 (n_I-1)^2+\frac{1}{5} \fc_2 q_I^2 \beta^4 \approx \fc_0 +\oc(\beta^2) \\
    a_{\beta \beta}& = \frac{1}{15} I_0^2 \left({12\beta ^2} \left(5 \fc_0 m_I^2+\fc_2 q_I^2\right)+5 \fc_1 (n_I-1)^2\right) \approx \frac{\fc_1 I_0^2 (n_I-1)^2}{3} + \oc(\beta^2) \\
    a_{I_0\beta} &= a_{\beta I_0} = \frac{1}{15} \beta  I_0 \left(30 \fc_0 {m_I} \left(m_I \beta^2+1\right)+5 \fc_1 (n_I-1)^2+6 \fc_2 {q_I^2}{\beta ^2}\right) \\ &\approx \beta I_0\left({2\fc_0 m_I}  + \frac{\fc_1 (n_I-1)^2}{3}\right) + \oc(\beta^3)\,,\\
    a_{\theta \theta} &= \frac{1}{15} \beta ^2 I_0^2 \left(5 \fc_1 (n_I-1)^2+9\fc_2 {q_I^2}{\beta ^2}\right) \approx \frac{\fc_1 I_0^2 (n_I-1)^2}{3}\beta^2  + \oc(\beta^4) \\
    a_{\phi \phi} &= \frac{1}{15} \beta ^2 I_0^2 \sin^2 \theta  \left(5 \fc_1 (n_I-1)^2+9\fc_2 {q_I^2}{\beta ^2}\right) \approx \frac{\fc_1 I_0^2 (n_I-1)^2}{3}\beta^2\sin^2\theta  + \oc(\beta^4) 
    \,,
\end{align}
Incorporating all higher multipoles ($\ell\geq 1$) of the anisotropy induced by kinematic effects, i.e.
\begin{align}
    I(\hn) = \sum_{\ell \geq 1}I_\ell \,\pc_\ell(\hn\cdot\hv{v}) =  4\pi\sum_{\ell \geq 2}\sum_{m=-\ell}^{\ell} \frac{I_\ell}{(2\ell+1)}\,Y_{\ell m}(\hn)Y_{\ell m}^{*}(\hv{v}) \,,
\end{align}
we obtain a Fisher matrix of the form
\begin{align}
    F_{ij} = \frac{2T\Delta f}{(4\pi f \sigma)^4}N_{\rm pair} \times \left[ \fc_0\frac{\partial I_0(1+m_I\beta^2)}{\partial \Theta_i}\frac{\partial I_0(1+m_I\beta^2)}{\partial \Theta_j} + 4\pi\sum_{\ell\geq 1}\sum_{m=-\ell}^{\ell}\frac{\fc_\ell}{(2\ell+1)^2}\frac{\partial I_{\ell m}}{\partial \Theta_i}\frac{\partial I_{\ell m}^{*}}{\partial \Theta_j} \right]\,,
\end{align}
where we defined $I_{\ell m} \equiv I_{\ell}\,Y_{\ell m}(\hv{v})$.

\end{appendix}

{\small



\providecommand{\href}[2]{#2}\begingroup\raggedright\endgroup

}

\end{document}